\documentclass[aps,prl,twocolumn,floatfix,english,showpacs,10pt,superscriptaddress]{revtex4-2}%
\usepackage{bm}
\usepackage{url}
\usepackage{hyperref}
\usepackage{amsmath,amssymb}
\usepackage{amsfonts}

\usepackage{graphicx} 
\usepackage{amsmath} 
\usepackage{hyperref} 
\usepackage{xcolor}
\usepackage{color}
\begin{document}

\title{Inverse Design of Winding Tuple for Non-Hermitian Topological Edge Modes}

\author{Zihe Yang}
\thanks{These authors contribute equally to this work.}
\affiliation{School of Physics, Huazhong University of Science and Technology, Wuhan 430074, P. R. China}

\author{Kunling Zhou}
\thanks{These authors contribute equally to this work.}
\affiliation{School of Physics, Huazhong University of Science and Technology, Wuhan 430074, P. R. China}

\author{Bowen Zeng}
\email[]{zengbowen@csust.edu.cn}
\affiliation{Hunan Provincial Key Laboratory of Flexible Electronic Materials Genome Engineering,
School of Physics and Electronic Sciences, Changsha University of Science and Technology, Changsha 410114, P. R. China}
\author{Yong Hu}
\email[]{huyong@hust.edu.cn}
\affiliation{School of Physics, Huazhong University of Science and Technology, Wuhan 430074, P. R. China}

\date{\today}

\begin{abstract}

The interplay between topological localization and non-Hermiticity localization in non-Hermitian crystal systems results in a diversity of shapes of topological edge modes (EMs), offering opportunities to manipulate these modes for potential topological applications. The characterization of the domain of EMs and the engineering of these EMs require detailed information about their wave functions, which conventional calculation of topological invariants cannot provide. In this Letter, by recognizing EMs as specified solutions of eigenequation, we derive their wave functions in an extended non-Hermitian Su-Schrieffer-Heeger model. We then inversely construct a winding tuple $\left \{ w_{\scriptscriptstyle  GBZ},w_{\scriptscriptstyle  BZ}\right \} $ that characterizes the existence of EMs and their spatial distribution. Moreover, we define a novel spectral winding number equivalent to $w_{\scriptscriptstyle  BZ}$, which is determined by the product of energies of different bands. The inverse design of topological invariants allows us to categorize the localized nature of EMs even in systems lacking sublattice symmetry, which can facilitate the manipulation and utilization of EMs in the development of novel quantum materials and devices.

\end{abstract}

\maketitle

\textit{Introduction}---The topological edge modes (EMs) localized at the edge correspond to nontrivial bulk topological invariants performed on the Brillouin zone (BZ), a phenomenon known as the bulk-boundary correspondence (BBC), which is a central concept in the field of celebrated topological band theory~\cite{berry1984quantal,zak1989berry,hasan2010colloquium,qi2011topological,shen2012topological,bernevig2013topological,asboth2016short,RevModPhys.88.035005,gong2018topological,hassani2020topological}. In non-Hermitian open systems, where non-Hermiticity arises from interactions with the environment, the conventional BBC may break down due to the systems' sensitivity to the boundary~\cite{ashida2020non,lee2016anomalous,kunst2018biorthogonal,xiong2018,yao2018edge,kawabata2019,rui2019topology,yokomizo2019non,PhysRevB.99.201103,zhang2020,borgnia2020non,okuma2020,yang2020non,rui2022non,rui2023hermitian,zeng2023radiation,cai2023edge,yu2024non,PhysRevLett.127.116801}. This sensitivity leads to the accumulation of macroscopic bulk states at the boundary under an open boundary condition (OBC) called the non-Hermitian skin effect (NHSE)~\cite{zhang2022review,okuma2020,longhi2019probing,song2019non,zhang2020,longhi2020unraveling,manna2023inner,lin2023topological,lei2024activating,yang2022designing} and equivalently reshapes the range of the allowed wave vector, the collection of which is referred to as the generalized Brillouin zone (GBZ)~\cite{yao2018edge,yokomizo2019non,yang2020non}. Note that the NHSE respects spectral topology, rather than band topology, which refers to the winding of the spectrum 
under a periodic boundary condition (PBC) with respect to the OBC spectrum defined on the GBZ~\cite{zhang2020,okuma2020,ding2022non}. Recently, Yao \textit{et al.}~\cite{yao2018edge} demonstrated that the bulk topological invariants performed on the GBZ can precisely feature the existence of EMs in the non-Hermitian systems, marking a reestablishment of BBC. 

In Hermitian topological systems, such as the one-dimensional Su-Schrieffer-Heeger (SSH) model with intercell coupling larger than intracell coupling~\cite{su1979solitons,heeger1988solitons}, pairs of EMs are separately distributed at the two ends of systems~\cite{asboth2016short}. In non-Hermitian systems, the interplay between topological localization and non-Hermiticity localization gives rise to diverse patterns of EMs~\cite{lee2016anomalous,longhi2018non,yin2018geometrical,rui2019pt,zhu2021delocalization,wang2022non,gao2020anomalous,wang2022extended,cheng2022competition,rui2023making,eek2024emergent,slootman2024breaking}. For example, the balance between these two localization mechanisms allows the delocalization of topological EMs, a phenomenon observed in experiments~\cite{wang2022non,gao2020anomalous}. This offers potential pathways for manipulating the EMs and designing an ``extended state in a localized continuum''~\cite{longhi2018non,zhu2021delocalization,wang2022non,gao2020anomalous,wang2022extended}. While in the NHSE-dominant region, two EMs are localized at the same ends together~\cite{cheng2022competition,hou2022deterministic,hou2023topological}. Several topological invariants, such as the winding number based on the evolution of pseudomagnetic fields or the spectrum, have been proposed for building the corresponding BBC~\cite{yin2018geometrical,zhu2021delocalization,yao2018edge,yang2020non,PhysRevLett.121.026808,PhysRevB.99.081302,RevModPhys.93.015005}, for the domain of EMs and these invariants appears to be consistent. However, this approach does not provide the wave function forms of EMs.
Additionally, the reason why such a BBC works well and the relationships between these topological invariants remain not entirely clear.

In this Letter, by considering EMs as special solutions distinct from bulk state solutions of the eigenequation, we derive the analytical solutions for the EMs in a generalized non-Hermitian SSH model. According to the existence condition and behavior of these analytical solutions, we inversely construct a topological winding tuple $\left \{ w_{\scriptscriptstyle  GBZ},w_{\scriptscriptstyle  BZ}\right \}$ for non-Hermitian topological EMs. A nontrivial $w_{\scriptscriptstyle  GBZ}$ defined on the GBZ characterizes the presence of EMs and $w_{\scriptscriptstyle  BZ}={0,\pm1}$ defined on the BZ corresponds to two EMs being separately distributed at two ends, or both localized together at the left (right) end, respectively. This winding tuple is exactly a combination of previously defined topological invariants from the literature. Meanwhile, this winding tuple is independent of symmetry and thereby is applicable to systems without sublattice symmetry (SLS). Topological invariants concerning the spectrum derived from $w_{\scriptscriptstyle  BZ}$ has the form of the product of energies of different bands, revealing the connection between energy subbands and potentially advancing the spectral winding number of multi-band systems. Our results provide a comprehensive understanding of topological EMs and the associated BBC in non-Hermitian systems.  
 
\begin{figure}[!ht]
\centering
\includegraphics[scale=0.16]{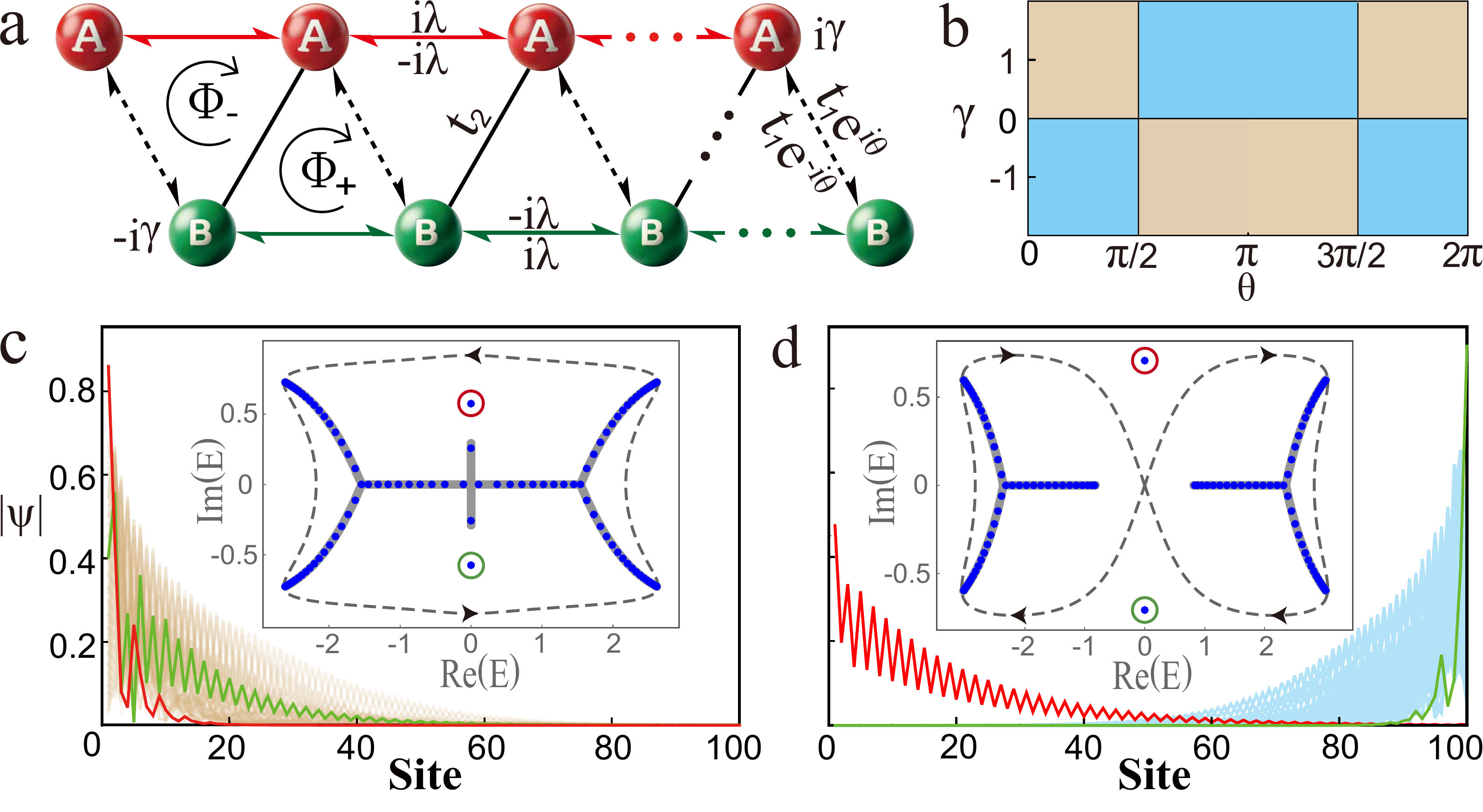}
\caption{(a) Non-Hermitian SSH model. (b) $ \gamma$-controlled and $\theta$-controlled NHSE, where the blue (beige) region indicates localized skin modes at the right (left) end. The wave function under the parameters (c)   \(t_2=1.4,\theta=0\) and (d)  \(t_2=2,\theta=\pi\), while keeping the same parameters \(\lambda=\gamma=t_1=1, N=50\), with the corresponding spectral winding by the PBC spectrum (gray dashed line) with respect to the OBC spectrum (blue points) shown in the insets. Here, the OBC spectrum in the thermodynamic limit is denoted by the gray solid lines. In (c) and (d), the energies and corresponding wave function of two EMs are marked by red (green) circles and lines, respectively.}
\label{fig.Model&Skin}
\end{figure}

\textit{Model}.---We consider a two-band non-Hermitian SSH model, as illustrated in Fig.~\ref{fig.Model&Skin}(a), with a Hamiltonian defined in momentum space,
\begin{equation}\label{M}
H(\beta)=\begin{pmatrix}
\mathrm{i} \gamma+\mathrm{i} \lambda \left(1/ \beta -\beta\right)  & t_1 e^{-\mathrm{i}\theta}+t_2 /\beta \\
t_1 e^{\mathrm{i}\theta}+t_2 \beta & -\mathrm{i} \gamma-\mathrm{i} \lambda \left( 1/\beta -\beta \right)
\end{pmatrix}.
\end{equation}
Here $H(\beta)$ represents the Bloch Hamiltonian for $\beta \in \text{BZ}$ or the non-Bloch Hamiltonian for $\beta \in \text{GBZ}$. Here, $\pm \mathrm{i} \gamma$ represents the on-site gain and loss at the $A/B$ sites, which are the only non-Hermitian terms in this model; $t_1 (t_2)$ and $\lambda$ are conjugated intracell (intercell) coupling and neighbor coupling between $A$-to-$A$ or $B$-to-$B$ sites. Without loss of generality, we only consider $\left\{t_1, t_2, \lambda\right\}\geq 0$. Additional phase $\pm \theta$ and $\pm\frac{\pi}{2}$ are introduced in $t_1$ and $\lambda$, respectively, which leads to magnetic fluxes of $\Phi_\pm=\frac{\pi}{2}\pm\theta$ in the model's triangular loops as shown in Fig.~\ref{fig.Model&Skin}(a). Such an SSH model enables gain (loss)-controlled and flux-controlled NHSE~\cite{yi2020non,li2022gain,wu2022flux} as shown in Fig.~\ref{fig.Model&Skin}(b), where the beige (blue) region represents the pile-up of bulk states at the left (right) boundary [see Fig.~\ref{fig.Model&Skin}(c-d)]. Such skewness corresponds to the positive (negative) winding of the PBC spectrum with respect to the OBC spectrum~\cite{zhang2022review,xue2021simple}, respectively, as shown in insets of Fig.~\ref{fig.Model&Skin}(c-d).

This model could support paired EMs, as shown in Fig.~\ref{fig.Model&Skin}(c-d), with their energy symmetrically distributed with respect to the real axis. The EMs can be either localized with skin modes in Fig.~\ref{fig.Model&Skin}(c) or one of them can be isolated at the left end as depicted in Fig.~\ref{fig.Model&Skin}(d). The positive spectral winding number for EMs and their localization in the same direction in Fig.~\ref{fig.Model&Skin}(c) appears to stick to conventional spectral BBC~\cite{zhang2020,yokomizo2020non,yang2020non}. However, for separately distributed EMs in Fig.~\ref{fig.Model&Skin}(d), the winding number is trivial.

It should be noted that establishing the BBC for EMs in our model may be challenging. One reason is that Eq.~\eqref{M} does not respect SLS and another reason is that the diagonal elements in Eq.~\eqref{M} depend on the wave vector. Thus, conventional approaches, such as defining the topological invariant using the ``$Q$ matrix'' or calculating the winding number of non-diagonal elements in the Hamiltonian, may not be applicable~\cite{chiu2016classification,yao2018edge}. Below, we attempt to establish the BBC for these EMs in non-Hermitian systems.

\textit{Analytic solutions for EMs}---We begin by analytically calculating the wave function of the edge modes. The conventional method for solving the system's wave function is to find a combination of the non-Bloch basis that satisfies the eigenequation including the bulk and boundary equations associated with the Hamiltonian matrix in real space~\cite{yao2018edge}, as also detailed in the Supplemental Material~\cite{SupplementalMaterials}. The trial wave functions at sites $A$ and $B$ in the $n$-th unit cell take the form of $\left( \phi_{Ai},\phi_{Bi} \right)^T\beta_i^n$ with $\beta_i$ denoting the non-Bloch wave vector. Substituting it into the system's bulk equations[(Eq.~(S2) in the Supplemental Material~\cite{SupplementalMaterials}], we find $\phi_{Ai} =\frac{t_1 e^{- \mathrm{i} \theta} + t_2/\beta_i}{E - q(\beta_i)} \phi_{Bi}$ and $\phi_{Bi} =\frac{t_1 e^{ \mathrm{i} \theta} + t_2\beta_i}{E + q(\beta_i)} \phi_{Ai}$, where $q(\beta_i)=\mathrm{i}\gamma +  \mathrm{i} \lambda(1/\beta_i -\beta_i)$. Using the relationship between $\phi_{Ai}$ and $\phi_{Bi}$, the characteristic equation between the wave vector and the eigenenergies can be obtained as
\begin{equation}
\begin{aligned}
    f(\beta,E) & =\lambda^2 \beta^2-(2 \gamma \lambda + t_1 t_2 e^{ -\mathrm{i} \theta}) \beta + (2 \gamma \lambda - t_1 t_2 e^{ \mathrm{i} \theta})/ \beta  \\
    & + \lambda^2/\beta^2+E^2 + \gamma^2 - 2\lambda^2 - t_1^2 - t_2^2= 0,
\label{fbe}
\end{aligned}
\end{equation}
which yields four roots for each energy $E$, denoted as $\{\beta_1,\beta_2,\beta_3,\beta_4\}$ sorted by their moduli. Then the wave function can be constructed as
\begin{equation}
\label{wave}
\begin{pmatrix}
\psi_A(n) \\
\psi_B(n)
\end{pmatrix}
= \sum_{i=1}^4
\begin{pmatrix}
\phi_{Ai} \\
\phi_{Bi}
\end{pmatrix}
\beta_i^n.
\end{equation}
Substituting Eq.~\eqref{wave} into the boundary equations [Eq.~(S3) in the Supplemental Material~\cite{SupplementalMaterials}] yields
\begin{equation}
\begin{cases}
 \sum_{i=1}^4 \mathrm{i} \lambda \alpha(\beta_i) \phi_{Bi} + t_2\phi_{Bi} = 0 ,\\
\sum_{i=1}^4-\mathrm{i} \lambda\phi_{Bi} = 0 ,\\
\sum_{i=1}^4-\mathrm{i} \lambda \alpha(\beta_i) \beta_i^{N+1} \phi_{Bi} = 0 ,\\
\sum_{i=1}^4t_2\alpha(\beta_i) \beta_i^{N+1} \phi_{Bi} +  \mathrm{i} \lambda\beta_i^{N+1} \phi_{Bi} = 0,
\end{cases}
\end{equation}
where $\alpha \left(\beta_i\right)= \frac{t_1 e^{- \mathrm{i} \theta} + t_2/\beta_i}{E - q(\beta_i)}$.
This equation can be further simplified as  \(M\phi=0\), where \(\phi=[\phi_{B1},\phi_{B2},\phi_{B3},\phi_{B4}]^{\mathsf{T}}\) and 
\begin{equation}
\setlength\arraycolsep{1.7pt}
{\small 
M=\begin{vmatrix}
\alpha(\beta_1) & \alpha(\beta_2) & \alpha(\beta_3) & \alpha(\beta_4) \\
1 & 1 & 1 & 1 \\
\alpha(\beta_1) \beta_1^{N+1} & \alpha(\beta_2) \beta_2^{N+1} & \alpha(\beta_3) \beta_3^{N+1} & \alpha(\beta_4) \beta_4^{N+1} \\
\beta_1^{N+1} & \beta_2^{N+1} & \beta_3^{N+1} & \beta_4^{N+1}
\end{vmatrix}.
}
\end{equation}

The solutions exist when det$ (M)=0$. For bulk states, this condition becomes the roots of the characteristic function fulfilling the GBZ condition \(|\beta_2|=|\beta_3|\) in the thermodynamic limit~\cite{yokomizo2019non}. The existence of EMs implies that special solutions can be found, which satisfy det$ (M)=0$ but with \(|\beta_2| < |\beta_3|\). By dividing the third and fourth rows of \(M\) by \(\beta_3^{N+1}\) and taking the thermodynamic limit \(N\to\infty\), \(M\) becomes a block upper triangular matrix
\begin{equation}
M=\begin{vmatrix}M_{I} & M_{III} \\
\textbf{0} & M_{IV}\end{vmatrix},
\end{equation}
where each element represents a $2 \times 2 $ submatrix. Now, \(\operatorname{det}(M)=\operatorname{det}(M_I)\times \operatorname{det}(M_{IV})\). The existence of special solutions requires either \(\operatorname{det}(M_I)\) or \(\operatorname{det}(M_{IV})\) to be zero. Let $\alpha \left(\beta_1\right) = \alpha \left(\beta_2\right)$ to ensure $\operatorname{det}(M_I)=0$. Obviously, $\beta_{1,2}$ can be considered as two roots of $C_{e1}=\alpha(\beta)$, where $C_{e1}$ is a constant and the subscript $e1$ denotes one EM if special solutions exist. Then we have the characteristic equation   
\begin{equation}
\begin{aligned}
f_{e1}(\beta,E_{e1}) =& - \mathrm{i} \lambda C_{e1} \beta + (t_2 +  \mathrm{i} \lambda C_{e1})/\beta\\
&+ (t_1 e^{- \mathrm{i} \theta} + \mathrm{i} \gamma C_{e1} - C_{e1} E_{e1})=0.
\end{aligned}
\label{edgeequation}
\end{equation}
Here, $E_{e1}$ is the energy of this EM. According to the spectral symmetry inherent in Eq.~\eqref{fbe}, the characteristic equation becomes $f_{e2}(\beta,E_{e2})=0$ for EM $e2$ with $E_{e2}=-E_{e1}$ and $\alpha \left(\beta_3\right) = \alpha \left(\beta_4\right) = C_{e2}$. The same results would be obtained if we first let $\operatorname{det}(M_{IV})=0$. By simultaneously solving the eigenequation of Eqs.~\eqref{M} and~\eqref{edgeequation} and eliminating $E$, the characteristic equation for EMs becomes 
\begin{equation}
\begin{aligned}
\Gamma_{e}\left(\beta \right) =& (-t_2C_{e}^2-2 \mathrm{i} \lambda C_{e})\beta+(2 \mathrm{i} \lambda C_{e}+t_2)/\beta\\
&+(-t_1 e^{\mathrm{i}\theta} C_{e}^2+t_1e^{-\mathrm{i}\theta}+2 \mathrm{i} \gamma C_{e})=0,
\end{aligned}
\label{Gamma}
\end{equation}
with index $e=\{e1;e2\}$. Here, $C_{e}^2 - i t_2 /\lambda C_{e} + 1=0$ (details available in the Supplemental Material~\cite{SupplementalMaterials}). With four roots solved, we further obtain the energies of EMs as

\begin{equation}
        E_e=\pm \frac{\gamma t_2/\lambda+2\mathrm{i}t_1\sin{\theta}}{C_{e1}-C_{e2}},
        \label{E_e}
\end{equation}
and the wave function for two EMs on each site 
\begin{equation}
\begin{aligned}
&\psi_{e1}(n) = \begin{pmatrix}
C_{e1} \beta_1^n - C_{e1} \beta_2^n \\
\beta_1^n - \beta_2^n
\end{pmatrix}\\
&\psi_{e2}(n) = \begin{pmatrix}
C_{e2} \beta_3^{n-(N+1)} - C_{e2} \beta_4^{n-(N+1)} \\
\beta_3^{n-(N+1)} - \beta_4^{n-(N+1)}
\end{pmatrix}.
\end{aligned}
\label{wavefunction}
\end{equation}
It can be observed that the $\psi_{e1}(n)$ and $\psi_{e2}(n)$ are predominantly governed by $\beta_2$ and $\beta_3$, respectively.

\begin{figure}[!ht]
\centering
\includegraphics[scale=0.185]{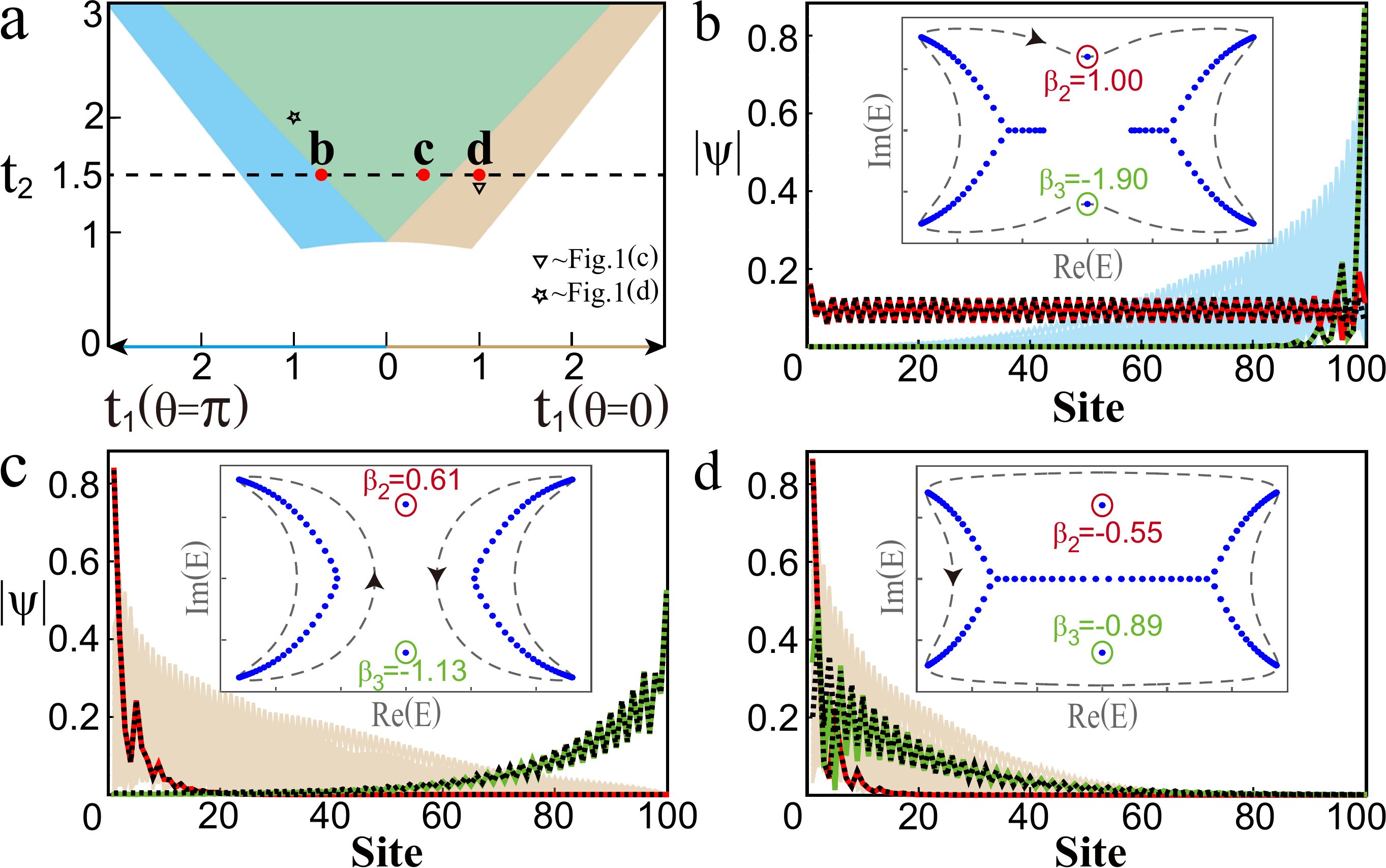}
\caption{(a) Topological phase diagram in the $(t_1,t_2)$ plane, where the white (colored) area represents the absence (presence) of EMs, and the blue (beige) and green region denotes two right-localized (left-localized) EMs and separately distributed EMs. Under fixed $t_2$ [dashed line in (a)] and other parameters $\lambda=\gamma=1$ and $N=50$, adjusting $t_1$ and $\theta$ leads to the transition between different regions. This transition is exemplified by the wave function of EMs for (b) $t_1=0.7, \theta = \pi$ [point \textbf{b} in (a)], (c) $t_1 = 0.4, \theta = 0$ [point \textbf{c} in (a)] and (d) $t_1 = 1, \theta = 0$ [point \textbf{d} in (a)]. In (b)-(d), the numerical calculations of wave function of EMs (red and green solid lines) are consistent with the theoretical results (black dashed line). The insets show the corresponding spectrum and $\beta_{2,3}$.}
\label{Phase}
\end{figure}
For arbitrary given parameters, the EMs exist if both two roots associated with a specified $E_e$ among four roots are larger (less) than the other two roots (also see Fig.~S1 in the Supplemental Material~\cite{SupplementalMaterials}). This allows us to map the system's phase diagram for EMs, as shown in Fig.~\ref{Phase}(a), where the white (colored) region represents the absence (presence) of EMs. From the wave function Eq.~\eqref{wavefunction}, it can be predicted that two EMs are localized at the left end when $\left|\beta_2\right|<\left|\beta_3\right|<1$ and the right end when $1<\left|\beta_2\right|<\left|\beta_3\right|$, and both ends when $\left|\beta_2\right|<1<\left|\beta_3\right|$, which corresponds to the beige, blue, and green region in Fig.~\ref{Phase}(a), respectively. These predictions are further verified and exemplified by numerical results in Figs.~\ref{Phase}(b)-~\ref{Phase}(d). The emergence of delocalized EM when $\left|\beta_2\right|=1$ or $\left|\beta_3\right|=1$, corresponding to a critical case between different colored regions such as point $\textbf{b}$ in Fig.~\ref{Phase}(a). An extended EM over the bulk is shown in Fig.~\ref{Phase}(b) with the associated energy located on the PBC spectrum. Thus, according to the root distribution but without the knowledge of any topological invariants, we derive the conditions for the existence of EMs and their spatial distribution. 

\textit{Inverse Design of Topological Winding Tuple}---
To establish the BBC for topological EMs, we inversely design the corresponding topological invariants by the root distribution as follows. Note that the existence of EMs requires the two largest (smallest) roots associated with $E_{e2}$ ($E_{e1}$) among four roots. According to the GBZ theory, for an energy in our model that does not belong to the OBC bulk spectrum, two roots must lie within the GBZ, while the other two roots must lie outside the GBZ~\cite{yokomizo2019non,yang2020non,yu2021generalized}. In other words, the GBZ bisects the four roots. Considering that $\Gamma_{e}(\beta)$ has a first-order pole at $\beta=0$, we can construct the winding number by $\Gamma_{e1;e2}(\beta)$ [only the solutions for EMs satisfy $\Gamma_{e1;e2}(\beta)=0$] along the GBZ
\begin{equation}
w_{e1;e2} = \oint_{ GBZ} \frac{1}{2\pi \mathrm{i}} \mathrm{d} \ln \Gamma_{e1;e2}(\beta).
\end{equation}
For the existence of EMs, $w_{e1}=1$ and $w_{e2}=-1$. A trivial $w_{e1}=0$ implies that $\beta_1$ and one of $\beta_3$ and $\beta_4$ are associated with $E_{e1}$. In this case, the corresponding $w_{e2}=0$, indicates the absence of EMs. Both nontrivial $w_{e1;e2}$ feature the existence of EMs, as does their combination in the form of 
\begin{equation}\label{wgbz}
w_{\scriptscriptstyle  GBZ} = \frac{w_{e1}-w_{e2}}{2}. 
\end{equation}

Given the dominance of $|\beta_2|$ and $|\beta_3|$ in Eq.~\eqref{wavefunction}, we naturally compare their magnitudes with 1 to determine the domain of EMs. Note that the BZ on the complex plane is a unit circle with a modulus of 1, which inspires us to define a winding number along the BZ
\begin{equation}
 w_{e1;e2}' = \oint_{ BZ} \frac{1}{2\pi \mathrm{i}} \mathrm{d} \ln \Gamma_{e1;e2}(\beta).
\label{w1}
\end{equation}
Here, $w_{e1}'=1(0)$ implies that there are two (one) roots for $E_{e1}$ in BZ and corresponds to this EM at the left (right) end. For the other EM, a similar distribution is characterized by $w_{e2}'=0(-1)$. The distribution of these two EMs can also be characterized by   
\begin{equation}
w_{\scriptscriptstyle  BZ} = w_{e1}' + w_{e2}',
\label{wbz}
\end{equation}
where $w_{_{BZ}}\in\{-1, 0, 1\}$. When $w_{_{BZ}} = 1(-1)$, three (one)  of the four $\beta_i$ of the EMs are within BZ, implying 
$|\beta_{2;3}| < 1$ ($|\beta_{2;3}| > 1$) and EMs are localized at the left end (right end). When $w_1 = 0$, $|\beta_2| < 1 < |\beta_3|$, two EMs are localized at opposite ends, a distribution analogous to the Hermitian scenario.

\begin{figure}[!ht]
\centering
\includegraphics[scale=0.226]{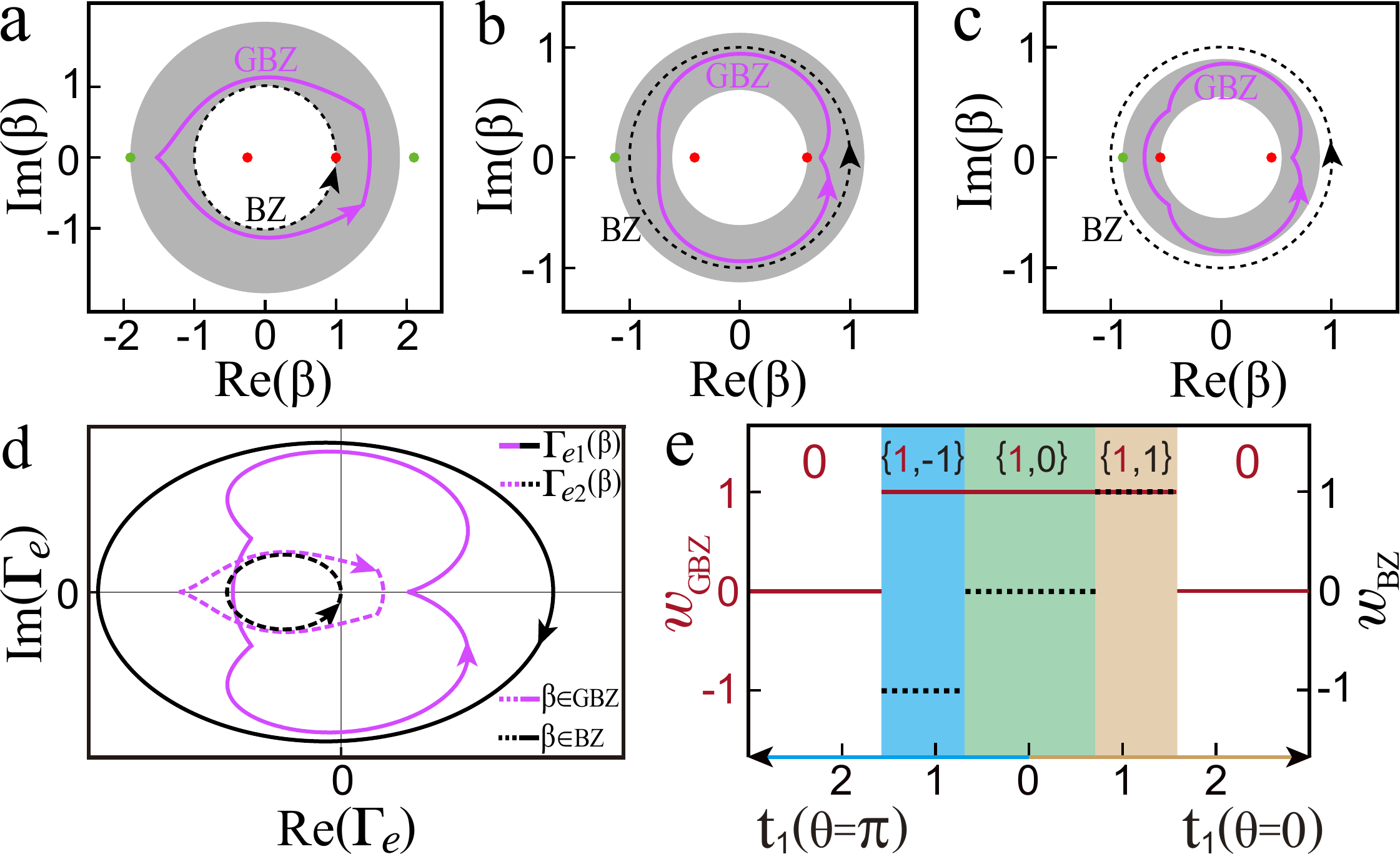}
\caption{The winding of GBZ (purple line) and BZ (black dashed line) relative to four roots under the same parameter used in (a) point \textbf{b} in Fig.~\ref{Phase}(a), (b) point \textbf{c} in Fig.~\ref{Phase}(a) and (c) point \textbf{d} in Fig.~\ref{Phase}(a). Here, among four roots, two smaller (larger) roots in moduli are denoted by red (green) points. For one root located on the BZ in (a), (d) shows that the evolution of $\Gamma_{e1;e2}(\beta)$ along the BZ and GBZ, and $\Gamma_{e2}(\beta \in \text{BZ})$ crosses the zero point. (e) The variation of defined winding tuple with $t_1$ capture the topological transition in  Fig.~\ref{Phase}(a).}
\label{GBZRoot}
\end{figure}

Thus, the topological EMs in non-Hermitian systems correspond to a winding tuple
\begin{equation}\label{wturple}
W = \left \{ w_{\scriptscriptstyle  GBZ},w_{\scriptscriptstyle  BZ}\right \}.
\end{equation}
The first element defined on the GBZ determines the existence of EMs, while the second element defined on the BZ determines the localized nature of the EMs. 

Such a winding tuple is consistent with the phase diagram in Fig.~\ref{Phase}(a), as illustrated in Fig.~\ref{GBZRoot}. Figs.~\ref{GBZRoot}(a)-~\ref{GBZRoot}(c) show four roots of EMs with respect to GBZ and BZ for three representative cases $\textbf{b}, \textbf{c},  \textbf{d}$ in Fig.~\ref{Phase}(a), corresponding to $w_{\scriptscriptstyle  GBZ}=1$, $W=\{1,1\}$, $W=\{1,-1\}$.  For a specified case $\textbf{b}$ in Fig.~\ref{Phase}(a), $w_{\scriptscriptstyle  BZ}$ is not well defined as the evolution of $\Gamma_{e2}(\beta)$ crosses the zero point as shown in Fig.~\ref{GBZRoot}(d). The phase transition in Fig.~\ref{Phase}(a) by adjusting $t_1e^{\mathrm{i}\theta}$ under fixed $t_2$ is captured by the variation of winding tuple as shown in Fig.~\ref{GBZRoot}(e). 

\textit{Cross-validation for topology}---In this section, we further address the relationship between winding tuple Eq.~\eqref{wturple} and previously defined topological invariants~\cite{yao2018edge,yin2018geometrical,yang2020non,zhu2021delocalization}. The wave function of single EM in systems with SLS typically is located at the single site $A$ or $B$. However, the wave function in Eq.~\eqref{wavefunction} is distributed on both sites, which inspires us to perform a similarity transformation $H^{\prime}(\beta)=U^{-1}H(\beta)U$ with transformation matrix \(U=\frac{1}{\mu}\begin{pmatrix}
d_- & d_+\\
1 & 1 
\end{pmatrix}.\) Here, $\mu=\sqrt{d_--d_+}$ is the normalization factor and $d_-(d_+)$ is the ratio of the amplitude between different sites. The new Hamiltonian reads 
\begin{equation}
    H^{\prime}(\beta) = \frac{1}{\mu^2}\begin{pmatrix}
m_0 + m(\beta) & R_+(\beta) \\
R_-(\beta) & -m_0 -m(\beta)
\end{pmatrix},
\end{equation}
with the diagonal element including a constant mass term 
\begin{equation}
    m_0=\mathrm{i} \gamma\left(d_++d_-\right)-d_+ d_- t_1 e^{\mathrm{i} \theta}+t_1 e^{-\mathrm{i} \theta}, 
\end{equation} a $\beta$-dependent term 
\begin{equation}
    m(\beta) = \mathrm{i}\lambda \left(d_+ + d_-\right)\left(1/\beta-\beta\right) -t_2 \beta + d_+d_-t_2 /\beta,
\end{equation} 
and the non-diagonal term 
\begin{equation}
\begin{aligned}
R_\pm(\beta)=& \pm  
\left[\left(-t_2 d_\pm^2-2 \mathrm{i}\lambda d_\pm\right) \beta + \left(2 \mathrm{i}\lambda d_\pm +t_2\right)/\beta \right.\\
& \left.+\left(-t_1 e^{\mathrm{i} \theta}d_\pm^2 +2\mathrm{i}\gamma d_\pm+t_1 e^{-\mathrm{i} \theta}\right)\right]
.
\end{aligned}
\end{equation}

To compare with the well-studied systems that have SLS, the selection of parameters should ensure the on-site mass term independent of $\beta$ [$m(\beta)=0$], which requires $d_+ + d_- = it_2/\lambda$ and $d_+ d_- = 1$. These conditions align with solutions for $C_{e}$, i.e., $\{d_+,d_-\}=\{C_{e2},C_{e1}\}$ and thereby $H^{\prime}\left( \beta\right)$ can be rewritten as 
\begin{equation}
    H^{\prime}(\beta) = \frac{1}{\mu^2}\begin{pmatrix}
m_0  & \Gamma_{e 2}(\beta) \\
-\Gamma_{e 1}(\beta) & -m_0 
\end{pmatrix}
\label{newH}
\end{equation}
with $m_0^2/\mu^4 = E_{e}^2$. Since the constant on-site mass term only alters the energy of EMs but not the wave function~\cite{kawarabayashi2021bulk}, the previously defined winding number on the GBZ~\cite{yao2018edge,yang2020non} and on the BZ~\cite{yin2018geometrical,zhu2021delocalization} returns to Eqs.~\ref{wgbz}
and~\ref{wbz}, respectively. These results also cross-validate our proposed analytic solutions for EMs. 

It should be noted, our model 
does not preserve those symmetries used for symmetry classification and belong to the $A$ class~\cite{kawabata2019}. However, the EMs in our model are not protected by concrete symmetry but are still protected by topological invariants $w_{\scriptscriptstyle  GBZ}$ and $w_{\scriptscriptstyle  BZ}$. These topological invariants are independent of symmetry, as shown below. For models similar to our work, Liang \textit{et al.} proposed that the topological phase transition can be characterized by a global Berry phase without symmetry breaking~\cite{liang2013topological}. Such a global Berry phase is equivalent to $w_{\scriptscriptstyle  GBZ}$ in our work, as detailed in the Supplemental Material~\cite{SupplementalMaterials}. This result shows that the topological invariant $w_{\scriptscriptstyle  GBZ}$ does not rely on symmetry.

\begin{figure}[!ht]
\centering
\includegraphics[scale=0.23]{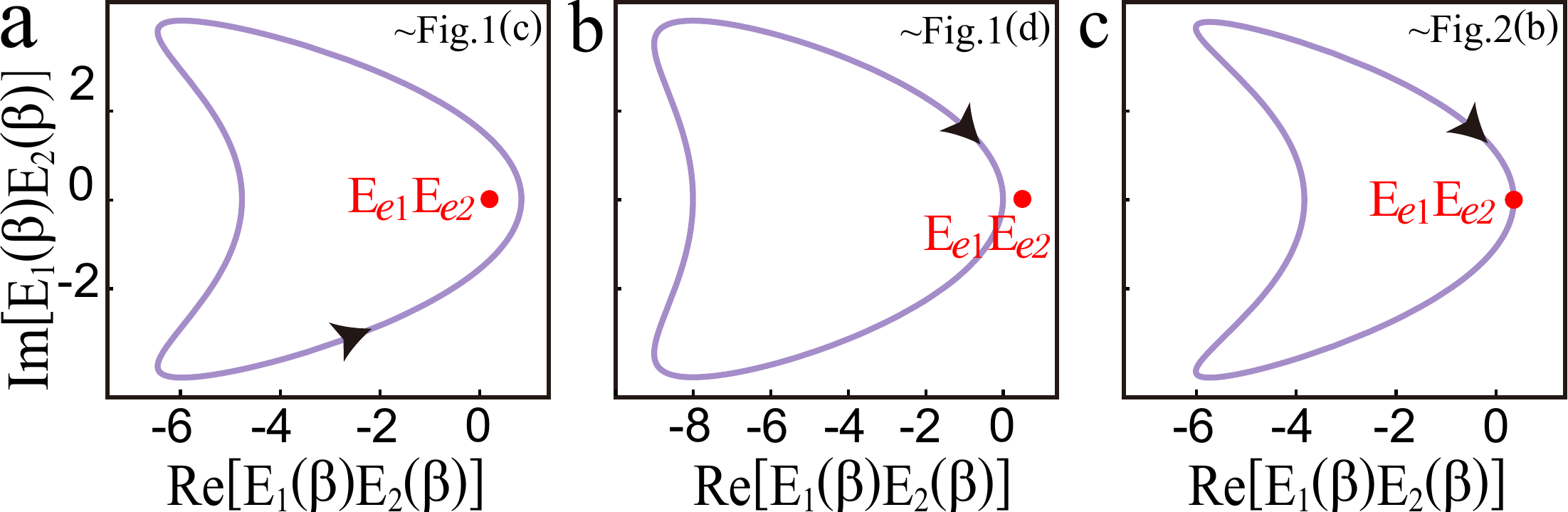}
 \caption{ As $\beta$ evolves on the BZ, the evolution of $\prod\limits E_i(\beta)$ under the same parameters used in (a) Fig.~\ref{fig.Model&Skin}(c), 
  (b) Fig.~\ref{fig.Model&Skin}(d) and (c) Fig.~\ref{Phase}(b). The product of energies of EMs $E_{e1}E_{e2}$ is denoted by a red dot.} 
\label{bzwinding}
\end{figure}

Combined with Eq.~\eqref{newH}, the winding number 
$ w_{\scriptscriptstyle  BZ}$ can be rewritten as  
\begin{equation}
\begin{aligned}
      w_{\scriptscriptstyle  BZ}=&\oint_{ BZ} \frac{1}{2\pi \mathrm{i}} \mathrm{d} \ln \Gamma_{e1}(\beta)\Gamma_{e2}(\beta)\\
     =&\oint_{ BZ} \frac{\mathrm{d} \ln [\operatorname{det}(H^{\prime})+m_0^2/\mu^4]+\mathrm{d}\ln(\mu^4)}{2\pi \mathrm{i}} .
\label{spctral1}
\end{aligned}
\end{equation}
The similarity transformation does not alter the eigenvalues, ensuring that \(\operatorname{det}(H^{\prime})=\operatorname{det}(H)=\prod_i E_i(\beta)\), where $E_i(\beta)$ represents the energy of the $i$-th band. Note that the diagonal elements of $H^{\prime}$ correspond to the eigenenergies of the edge modes, i.e., $m_0^2/\mu^4 = E_{e}^2=-\prod_i E_{ei}$, and the integral of a constant $\mu^4$ over the loop equals zero. We can define a new topological invariant by the spectrum 
\begin{equation}\label{bze}
    w_{\scriptscriptstyle  BZ}(E) =\frac{1}{2\pi \mathrm{i}}\oint_{ BZ} \mathrm{d} \ln \left[\prod\limits_{i=1;2} E_i(\beta)-\prod\limits_{i=1;2} E_{ei}\right],
\end{equation}
which is equivalent to $w_{\scriptscriptstyle  BZ}$. Different from a conventional spectral winding number~\cite{yang2020non} that refers to the winding of the PBC spectrum related to the reference energy, $w_{\scriptscriptstyle  BZ}(E)$ takes the form of the product of energies of different bands. However, they also share a similarity, namely that the topological invariant based on the spectrum typically does not depend on symmetry~\cite{yang2020non}. Thus, $w_{\scriptscriptstyle  BZ}$ is also independent of symmetry.

As shown in Fig.~\ref{bzwinding}, $w_{\scriptscriptstyle  BZ}(E)$ fits well with the previous calculations. 
Two left-localized EMs in Fig.~\ref{fig.Model&Skin} (c) and separately distributed EMs in Fig.~\ref{fig.Model&Skin} (d) are captured by $w_{\scriptscriptstyle  BZ}(E)=1$ and $w_{\scriptscriptstyle  BZ}(E)=0$.  A critical case is shown in Fig.~\ref{bzwinding}(c) where the product of two EMs falls in the $\prod\limits E_i(\beta)$, corresponding to a delocalized state in  Fig.~\ref{Phase}(b).

\textit{Conclusions}---In this Letter, we derive the analytic solutions for EMs and inversely construct the topological winding tuple $W = \left \{ w_{\scriptscriptstyle  GBZ},w_{\scriptscriptstyle  BZ}\right \}$, corresponding to the existence and localized nature of EMs, respectively.  We also define a spectral winding number based on the product of the energy of different bands, offering a different perspective for unveiling the extraordinary EMs in non-Hermitian systems. Our study not only advances the theoretical understanding of non-Hermitian topological EMs but also suggests potential avenues for manipulating and utilizing EMs for future experimental explorations aimed at harnessing the practical applications of topological phenomena. 

We thank Wenbin Rui, Xiaoshang Jin, and Liangliang Wan for helpful discussions. This work is supported by the
Natural Science Foundation of Hunan Province (2024JJ6011) and Innovation Program for Quantum
Science and Technology (Grant No. 2021ZD0302300).

\bibliography{manuscript.bib}

\begin{thebibliography}{66}%
\makeatletter
\providecommand \@ifxundefined [1]{%
 \@ifx{#1\undefined}
}%
\providecommand \@ifnum [1]{%
 \ifnum #1\expandafter \@firstoftwo
 \else \expandafter \@secondoftwo
 \fi
}%
\providecommand \@ifx [1]{%
 \ifx #1\expandafter \@firstoftwo
 \else \expandafter \@secondoftwo
 \fi
}%
\providecommand \natexlab [1]{#1}%
\providecommand \enquote  [1]{``#1''}%
\providecommand \bibnamefont  [1]{#1}%
\providecommand \bibfnamefont [1]{#1}%
\providecommand \citenamefont [1]{#1}%
\providecommand \href@noop [0]{\@secondoftwo}%
\providecommand \href [0]{\begingroup \@sanitize@url \@href}%
\providecommand \@href[1]{\@@startlink{#1}\@@href}%
\providecommand \@@href[1]{\endgroup#1\@@endlink}%
\providecommand \@sanitize@url [0]{\catcode `\\12\catcode `\$12\catcode `\&12\catcode `\#12\catcode `\^12\catcode `\_12\catcode `\%12\relax}%
\providecommand \@@startlink[1]{}%
\providecommand \@@endlink[0]{}%
\providecommand \url  [0]{\begingroup\@sanitize@url \@url }%
\providecommand \@url [1]{\endgroup\@href {#1}{\urlprefix }}%
\providecommand \urlprefix  [0]{URL }%
\providecommand \Eprint [0]{\href }%
\providecommand \doibase [0]{https://doi.org/}%
\providecommand \selectlanguage [0]{\@gobble}%
\providecommand \bibinfo  [0]{\@secondoftwo}%
\providecommand \bibfield  [0]{\@secondoftwo}%
\providecommand \translation [1]{[#1]}%
\providecommand \BibitemOpen [0]{}%
\providecommand \bibitemStop [0]{}%
\providecommand \bibitemNoStop [0]{.\EOS\space}%
\providecommand \EOS [0]{\spacefactor3000\relax}%
\providecommand \BibitemShut  [1]{\csname bibitem#1\endcsname}%
\let\auto@bib@innerbib\@empty
\bibitem [{\citenamefont {Berry}(1984)}]{berry1984quantal}%
  \BibitemOpen
  \bibfield  {author} {\bibinfo {author} {\bibfnamefont {M.~V.}\ \bibnamefont {Berry}},\ }\bibfield  {title} {\bibinfo {title} {Quantal phase factors accompanying adiabatic changes},\ }\href@noop {} {\bibfield  {journal} {\bibinfo  {journal} {Proc. R. Soc. London, Ser. A}\ }\textbf {\bibinfo {volume} {392}},\ \bibinfo {pages} {45} (\bibinfo {year} {1984})}\BibitemShut {NoStop}%
\bibitem [{\citenamefont {Zak}(1989)}]{zak1989berry}%
  \BibitemOpen
  \bibfield  {author} {\bibinfo {author} {\bibfnamefont {J.}~\bibnamefont {Zak}},\ }\bibfield  {title} {\bibinfo {title} {Berry’s phase for energy bands in solids},\ }\href@noop {} {\bibfield  {journal} {\bibinfo  {journal} {Phys. Rev. Lett.}\ }\textbf {\bibinfo {volume} {62}},\ \bibinfo {pages} {2747} (\bibinfo {year} {1989})}\BibitemShut {NoStop}%
\bibitem [{\citenamefont {Hasan}\ and\ \citenamefont {Kane}(2010)}]{hasan2010colloquium}%
  \BibitemOpen
  \bibfield  {author} {\bibinfo {author} {\bibfnamefont {M.~Z.}\ \bibnamefont {Hasan}}\ and\ \bibinfo {author} {\bibfnamefont {C.~L.}\ \bibnamefont {Kane}},\ }\bibfield  {title} {\bibinfo {title} {Colloquium: topological insulators},\ }\href@noop {} {\bibfield  {journal} {\bibinfo  {journal} {Reviews of modern physics}\ }\textbf {\bibinfo {volume} {82}},\ \bibinfo {pages} {3045} (\bibinfo {year} {2010})}\BibitemShut {NoStop}%
\bibitem [{\citenamefont {Qi}\ and\ \citenamefont {Zhang}(2011)}]{qi2011topological}%
  \BibitemOpen
  \bibfield  {author} {\bibinfo {author} {\bibfnamefont {X.-L.}\ \bibnamefont {Qi}}\ and\ \bibinfo {author} {\bibfnamefont {S.-C.}\ \bibnamefont {Zhang}},\ }\bibfield  {title} {\bibinfo {title} {Topological insulators and superconductors},\ }\href@noop {} {\bibfield  {journal} {\bibinfo  {journal} {Rev. Mod. Phys.}\ }\textbf {\bibinfo {volume} {83}},\ \bibinfo {pages} {1057} (\bibinfo {year} {2011})}\BibitemShut {NoStop}%
\bibitem [{\citenamefont {Shen}(2012)}]{shen2012topological}%
  \BibitemOpen
  \bibfield  {author} {\bibinfo {author} {\bibfnamefont {S.-Q.}\ \bibnamefont {Shen}},\ }\href@noop {} {\emph {\bibinfo {title} {Topological insulators}}},\ Vol.\ \bibinfo {volume} {174}\ (\bibinfo  {publisher} {Springer},\ \bibinfo {year} {2012})\BibitemShut {NoStop}%
\bibitem [{\citenamefont {Bernevig}(2013)}]{bernevig2013topological}%
  \BibitemOpen
  \bibfield  {author} {\bibinfo {author} {\bibfnamefont {B.~A.}\ \bibnamefont {Bernevig}},\ }\bibfield  {title} {\bibinfo {title} {Topological insulators and topological superconductors},\ }in\ \href@noop {} {\emph {\bibinfo {booktitle} {Topological Insulators and Topological Superconductors}}}\ (\bibinfo  {publisher} {Princeton university press},\ \bibinfo {year} {2013})\BibitemShut {NoStop}%
\bibitem [{\citenamefont {Asb{\'o}th}\ \emph {et~al.}(2016)\citenamefont {Asb{\'o}th}, \citenamefont {Oroszl{\'a}ny},\ and\ \citenamefont {P{\'a}lyi}}]{asboth2016short}%
  \BibitemOpen
  \bibfield  {author} {\bibinfo {author} {\bibfnamefont {J.~K.}\ \bibnamefont {Asb{\'o}th}}, \bibinfo {author} {\bibfnamefont {L.}~\bibnamefont {Oroszl{\'a}ny}},\ and\ \bibinfo {author} {\bibfnamefont {A.}~\bibnamefont {P{\'a}lyi}},\ }\bibfield  {title} {\bibinfo {title} {A short course on topological insulators},\ }\href@noop {} {\bibfield  {journal} {\bibinfo  {journal} {Lecture notes in physics}\ }\textbf {\bibinfo {volume} {919}} (\bibinfo {year} {2016})}\BibitemShut {NoStop}%
\bibitem [{\citenamefont {Chiu}\ \emph {et~al.}(2016{\natexlab{a}})\citenamefont {Chiu}, \citenamefont {Teo}, \citenamefont {Schnyder},\ and\ \citenamefont {Ryu}}]{RevModPhys.88.035005}%
  \BibitemOpen
  \bibfield  {author} {\bibinfo {author} {\bibfnamefont {C.-K.}\ \bibnamefont {Chiu}}, \bibinfo {author} {\bibfnamefont {J.~C.~Y.}\ \bibnamefont {Teo}}, \bibinfo {author} {\bibfnamefont {A.~P.}\ \bibnamefont {Schnyder}},\ and\ \bibinfo {author} {\bibfnamefont {S.}~\bibnamefont {Ryu}},\ }\bibfield  {title} {\bibinfo {title} {Classification of topological quantum matter with symmetries},\ }\href {https://doi.org/10.1103/RevModPhys.88.035005} {\bibfield  {journal} {\bibinfo  {journal} {Rev. Mod. Phys.}\ }\textbf {\bibinfo {volume} {88}},\ \bibinfo {pages} {035005} (\bibinfo {year} {2016}{\natexlab{a}})}\BibitemShut {NoStop}%
\bibitem [{\citenamefont {Gong}\ \emph {et~al.}(2018)\citenamefont {Gong}, \citenamefont {Ashida}, \citenamefont {Kawabata}, \citenamefont {Takasan}, \citenamefont {Higashikawa},\ and\ \citenamefont {Ueda}}]{gong2018topological}%
  \BibitemOpen
  \bibfield  {author} {\bibinfo {author} {\bibfnamefont {Z.}~\bibnamefont {Gong}}, \bibinfo {author} {\bibfnamefont {Y.}~\bibnamefont {Ashida}}, \bibinfo {author} {\bibfnamefont {K.}~\bibnamefont {Kawabata}}, \bibinfo {author} {\bibfnamefont {K.}~\bibnamefont {Takasan}}, \bibinfo {author} {\bibfnamefont {S.}~\bibnamefont {Higashikawa}},\ and\ \bibinfo {author} {\bibfnamefont {M.}~\bibnamefont {Ueda}},\ }\bibfield  {title} {\bibinfo {title} {Topological phases of non-hermitian systems},\ }\href@noop {} {\bibfield  {journal} {\bibinfo  {journal} {Physical Review X}\ }\textbf {\bibinfo {volume} {8}},\ \bibinfo {pages} {031079} (\bibinfo {year} {2018})}\BibitemShut {NoStop}%
\bibitem [{\citenamefont {Hassani~Gangaraj}\ \emph {et~al.}(2020)\citenamefont {Hassani~Gangaraj}, \citenamefont {Valagiannopoulos},\ and\ \citenamefont {Monticone}}]{hassani2020topological}%
  \BibitemOpen
  \bibfield  {author} {\bibinfo {author} {\bibfnamefont {S.~A.}\ \bibnamefont {Hassani~Gangaraj}}, \bibinfo {author} {\bibfnamefont {C.}~\bibnamefont {Valagiannopoulos}},\ and\ \bibinfo {author} {\bibfnamefont {F.}~\bibnamefont {Monticone}},\ }\bibfield  {title} {\bibinfo {title} {Topological scattering resonances at ultralow frequencies},\ }\href@noop {} {\bibfield  {journal} {\bibinfo  {journal} {Physical Review Research}\ }\textbf {\bibinfo {volume} {2}},\ \bibinfo {pages} {023180} (\bibinfo {year} {2020})}\BibitemShut {NoStop}%
\bibitem [{\citenamefont {Ashida}\ \emph {et~al.}(2020)\citenamefont {Ashida}, \citenamefont {Gong},\ and\ \citenamefont {Ueda}}]{ashida2020non}%
  \BibitemOpen
  \bibfield  {author} {\bibinfo {author} {\bibfnamefont {Y.}~\bibnamefont {Ashida}}, \bibinfo {author} {\bibfnamefont {Z.}~\bibnamefont {Gong}},\ and\ \bibinfo {author} {\bibfnamefont {M.}~\bibnamefont {Ueda}},\ }\bibfield  {title} {\bibinfo {title} {Non-hermitian physics},\ }\href@noop {} {\bibfield  {journal} {\bibinfo  {journal} {Advances in Physics}\ }\textbf {\bibinfo {volume} {69}},\ \bibinfo {pages} {249} (\bibinfo {year} {2020})}\BibitemShut {NoStop}%
\bibitem [{\citenamefont {Lee}(2016)}]{lee2016anomalous}%
  \BibitemOpen
  \bibfield  {author} {\bibinfo {author} {\bibfnamefont {T.~E.}\ \bibnamefont {Lee}},\ }\bibfield  {title} {\bibinfo {title} {Anomalous edge state in a non-hermitian lattice},\ }\href@noop {} {\bibfield  {journal} {\bibinfo  {journal} {Physical review letters}\ }\textbf {\bibinfo {volume} {116}},\ \bibinfo {pages} {133903} (\bibinfo {year} {2016})}\BibitemShut {NoStop}%
\bibitem [{\citenamefont {Kunst}\ \emph {et~al.}(2018{\natexlab{a}})\citenamefont {Kunst}, \citenamefont {Edvardsson}, \citenamefont {Budich},\ and\ \citenamefont {Bergholtz}}]{kunst2018biorthogonal}%
  \BibitemOpen
  \bibfield  {author} {\bibinfo {author} {\bibfnamefont {F.~K.}\ \bibnamefont {Kunst}}, \bibinfo {author} {\bibfnamefont {E.}~\bibnamefont {Edvardsson}}, \bibinfo {author} {\bibfnamefont {J.~C.}\ \bibnamefont {Budich}},\ and\ \bibinfo {author} {\bibfnamefont {E.~J.}\ \bibnamefont {Bergholtz}},\ }\bibfield  {title} {\bibinfo {title} {Biorthogonal bulk-boundary correspondence in non-hermitian systems},\ }\href@noop {} {\bibfield  {journal} {\bibinfo  {journal} {Phys. Rev. Lett.}\ }\textbf {\bibinfo {volume} {121}},\ \bibinfo {pages} {026808} (\bibinfo {year} {2018}{\natexlab{a}})}\BibitemShut {NoStop}%
\bibitem [{\citenamefont {Xiong}(2018)}]{xiong2018}%
  \BibitemOpen
  \bibfield  {author} {\bibinfo {author} {\bibfnamefont {Y.}~\bibnamefont {Xiong}},\ }\bibfield  {title} {\bibinfo {title} {Why does bulk boundary correspondence fail in some non-hermitian topological models},\ }\href@noop {} {\bibfield  {journal} {\bibinfo  {journal} {J. Phys. Commun.}\ }\textbf {\bibinfo {volume} {2}},\ \bibinfo {pages} {035043} (\bibinfo {year} {2018})}\BibitemShut {NoStop}%
\bibitem [{\citenamefont {Yao}\ and\ \citenamefont {Wang}(2018)}]{yao2018edge}%
  \BibitemOpen
  \bibfield  {author} {\bibinfo {author} {\bibfnamefont {S.}~\bibnamefont {Yao}}\ and\ \bibinfo {author} {\bibfnamefont {Z.}~\bibnamefont {Wang}},\ }\bibfield  {title} {\bibinfo {title} {Edge states and topological invariants of non-hermitian systems},\ }\href@noop {} {\bibfield  {journal} {\bibinfo  {journal} {Phys. Rev. Lett.}\ }\textbf {\bibinfo {volume} {121}},\ \bibinfo {pages} {086803} (\bibinfo {year} {2018})}\BibitemShut {NoStop}%
\bibitem [{\citenamefont {Kawabata}\ \emph {et~al.}(2019)\citenamefont {Kawabata}, \citenamefont {Shiozaki}, \citenamefont {Ueda},\ and\ \citenamefont {Sato}}]{kawabata2019}%
  \BibitemOpen
  \bibfield  {author} {\bibinfo {author} {\bibfnamefont {K.}~\bibnamefont {Kawabata}}, \bibinfo {author} {\bibfnamefont {K.}~\bibnamefont {Shiozaki}}, \bibinfo {author} {\bibfnamefont {M.}~\bibnamefont {Ueda}},\ and\ \bibinfo {author} {\bibfnamefont {M.}~\bibnamefont {Sato}},\ }\bibfield  {title} {\bibinfo {title} {Symmetry and topology in non-hermitian physics},\ }\href@noop {} {\bibfield  {journal} {\bibinfo  {journal} {Phys. Rev. X}\ }\textbf {\bibinfo {volume} {9}},\ \bibinfo {pages} {041015} (\bibinfo {year} {2019})}\BibitemShut {NoStop}%
\bibitem [{\citenamefont {Rui}\ \emph {et~al.}(2019{\natexlab{a}})\citenamefont {Rui}, \citenamefont {Zhao},\ and\ \citenamefont {Schnyder}}]{rui2019topology}%
  \BibitemOpen
  \bibfield  {author} {\bibinfo {author} {\bibfnamefont {W.}~\bibnamefont {Rui}}, \bibinfo {author} {\bibfnamefont {Y.}~\bibnamefont {Zhao}},\ and\ \bibinfo {author} {\bibfnamefont {A.~P.}\ \bibnamefont {Schnyder}},\ }\bibfield  {title} {\bibinfo {title} {Topology and exceptional points of massive dirac models with generic non-hermitian perturbations},\ }\href@noop {} {\bibfield  {journal} {\bibinfo  {journal} {Physical Review B}\ }\textbf {\bibinfo {volume} {99}},\ \bibinfo {pages} {241110} (\bibinfo {year} {2019}{\natexlab{a}})}\BibitemShut {NoStop}%
\bibitem [{\citenamefont {Yokomizo}\ and\ \citenamefont {Murakami}(2019)}]{yokomizo2019non}%
  \BibitemOpen
  \bibfield  {author} {\bibinfo {author} {\bibfnamefont {K.}~\bibnamefont {Yokomizo}}\ and\ \bibinfo {author} {\bibfnamefont {S.}~\bibnamefont {Murakami}},\ }\bibfield  {title} {\bibinfo {title} {Non-bloch band theory of non-hermitian systems},\ }\href@noop {} {\bibfield  {journal} {\bibinfo  {journal} {Phys. Rev. Lett.}\ }\textbf {\bibinfo {volume} {123}},\ \bibinfo {pages} {066404} (\bibinfo {year} {2019})}\BibitemShut {NoStop}%
\bibitem [{\citenamefont {Lee}\ and\ \citenamefont {Thomale}(2019)}]{PhysRevB.99.201103}%
  \BibitemOpen
  \bibfield  {author} {\bibinfo {author} {\bibfnamefont {C.~H.}\ \bibnamefont {Lee}}\ and\ \bibinfo {author} {\bibfnamefont {R.}~\bibnamefont {Thomale}},\ }\bibfield  {title} {\bibinfo {title} {Anatomy of skin modes and topology in non-hermitian systems},\ }\href {https://doi.org/10.1103/PhysRevB.99.201103} {\bibfield  {journal} {\bibinfo  {journal} {Phys. Rev. B}\ }\textbf {\bibinfo {volume} {99}},\ \bibinfo {pages} {201103(R)} (\bibinfo {year} {2019})}\BibitemShut {NoStop}%
\bibitem [{\citenamefont {Zhang}\ \emph {et~al.}(2020)\citenamefont {Zhang}, \citenamefont {Yang},\ and\ \citenamefont {Fang}}]{zhang2020}%
  \BibitemOpen
  \bibfield  {author} {\bibinfo {author} {\bibfnamefont {K.}~\bibnamefont {Zhang}}, \bibinfo {author} {\bibfnamefont {Z.}~\bibnamefont {Yang}},\ and\ \bibinfo {author} {\bibfnamefont {C.}~\bibnamefont {Fang}},\ }\bibfield  {title} {\bibinfo {title} {Correspondence between winding numbers and skin modes in non-hermitian systems},\ }\href@noop {} {\bibfield  {journal} {\bibinfo  {journal} {Phys. Rev. Lett.}\ }\textbf {\bibinfo {volume} {125}},\ \bibinfo {pages} {126402} (\bibinfo {year} {2020})}\BibitemShut {NoStop}%
\bibitem [{\citenamefont {Borgnia}\ \emph {et~al.}(2020)\citenamefont {Borgnia}, \citenamefont {Kruchkov},\ and\ \citenamefont {Slager}}]{borgnia2020non}%
  \BibitemOpen
  \bibfield  {author} {\bibinfo {author} {\bibfnamefont {D.~S.}\ \bibnamefont {Borgnia}}, \bibinfo {author} {\bibfnamefont {A.~J.}\ \bibnamefont {Kruchkov}},\ and\ \bibinfo {author} {\bibfnamefont {R.-J.}\ \bibnamefont {Slager}},\ }\bibfield  {title} {\bibinfo {title} {Non-hermitian boundary modes and topology},\ }\href@noop {} {\bibfield  {journal} {\bibinfo  {journal} {Phys. Rev. Lett.}\ }\textbf {\bibinfo {volume} {124}},\ \bibinfo {pages} {056802} (\bibinfo {year} {2020})}\BibitemShut {NoStop}%
\bibitem [{\citenamefont {Okuma}\ \emph {et~al.}(2020)\citenamefont {Okuma}, \citenamefont {Kawabata}, \citenamefont {Shiozaki},\ and\ \citenamefont {Sato}}]{okuma2020}%
  \BibitemOpen
  \bibfield  {author} {\bibinfo {author} {\bibfnamefont {N.}~\bibnamefont {Okuma}}, \bibinfo {author} {\bibfnamefont {K.}~\bibnamefont {Kawabata}}, \bibinfo {author} {\bibfnamefont {K.}~\bibnamefont {Shiozaki}},\ and\ \bibinfo {author} {\bibfnamefont {M.}~\bibnamefont {Sato}},\ }\bibfield  {title} {\bibinfo {title} {Topological origin of non-hermitian skin effects},\ }\href@noop {} {\bibfield  {journal} {\bibinfo  {journal} {Phys. Rev. Lett.}\ }\textbf {\bibinfo {volume} {124}},\ \bibinfo {pages} {086801} (\bibinfo {year} {2020})}\BibitemShut {NoStop}%
\bibitem [{\citenamefont {Yang}\ \emph {et~al.}(2020)\citenamefont {Yang}, \citenamefont {Zhang}, \citenamefont {Fang},\ and\ \citenamefont {Hu}}]{yang2020non}%
  \BibitemOpen
  \bibfield  {author} {\bibinfo {author} {\bibfnamefont {Z.}~\bibnamefont {Yang}}, \bibinfo {author} {\bibfnamefont {K.}~\bibnamefont {Zhang}}, \bibinfo {author} {\bibfnamefont {C.}~\bibnamefont {Fang}},\ and\ \bibinfo {author} {\bibfnamefont {J.}~\bibnamefont {Hu}},\ }\bibfield  {title} {\bibinfo {title} {Non-hermitian bulk-boundary correspondence and auxiliary generalized brillouin zone theory},\ }\href@noop {} {\bibfield  {journal} {\bibinfo  {journal} {Phys. Rev. Lett.}\ }\textbf {\bibinfo {volume} {125}},\ \bibinfo {pages} {226402} (\bibinfo {year} {2020})}\BibitemShut {NoStop}%
\bibitem [{\citenamefont {Rui}\ \emph {et~al.}(2022)\citenamefont {Rui}, \citenamefont {Zheng}, \citenamefont {Wang},\ and\ \citenamefont {Wang}}]{rui2022non}%
  \BibitemOpen
  \bibfield  {author} {\bibinfo {author} {\bibfnamefont {W.}~\bibnamefont {Rui}}, \bibinfo {author} {\bibfnamefont {Z.}~\bibnamefont {Zheng}}, \bibinfo {author} {\bibfnamefont {C.}~\bibnamefont {Wang}},\ and\ \bibinfo {author} {\bibfnamefont {Z.}~\bibnamefont {Wang}},\ }\bibfield  {title} {\bibinfo {title} {Non-hermitian spatial symmetries and their stabilized normal and exceptional topological semimetals},\ }\href@noop {} {\bibfield  {journal} {\bibinfo  {journal} {Physical Review Letters}\ }\textbf {\bibinfo {volume} {128}},\ \bibinfo {pages} {226401} (\bibinfo {year} {2022})}\BibitemShut {NoStop}%
\bibitem [{\citenamefont {Rui}\ \emph {et~al.}(2023{\natexlab{a}})\citenamefont {Rui}, \citenamefont {Zhao},\ and\ \citenamefont {Wang}}]{rui2023hermitian}%
  \BibitemOpen
  \bibfield  {author} {\bibinfo {author} {\bibfnamefont {W.}~\bibnamefont {Rui}}, \bibinfo {author} {\bibfnamefont {Y.}~\bibnamefont {Zhao}},\ and\ \bibinfo {author} {\bibfnamefont {Z.}~\bibnamefont {Wang}},\ }\bibfield  {title} {\bibinfo {title} {Hermitian topologies originating from non-hermitian braidings},\ }\href@noop {} {\bibfield  {journal} {\bibinfo  {journal} {Physical Review B}\ }\textbf {\bibinfo {volume} {108}},\ \bibinfo {pages} {165105} (\bibinfo {year} {2023}{\natexlab{a}})}\BibitemShut {NoStop}%
\bibitem [{\citenamefont {Zeng}\ and\ \citenamefont {Yu}(2023)}]{zeng2023radiation}%
  \BibitemOpen
  \bibfield  {author} {\bibinfo {author} {\bibfnamefont {B.}~\bibnamefont {Zeng}}\ and\ \bibinfo {author} {\bibfnamefont {T.}~\bibnamefont {Yu}},\ }\bibfield  {title} {\bibinfo {title} {Radiation-free and non-hermitian topology inertial defect states of on-chip magnons},\ }\href@noop {} {\bibfield  {journal} {\bibinfo  {journal} {Physical Review Research}\ }\textbf {\bibinfo {volume} {5}},\ \bibinfo {pages} {013003} (\bibinfo {year} {2023})}\BibitemShut {NoStop}%
\bibitem [{\citenamefont {Cai}\ \emph {et~al.}(2023)\citenamefont {Cai}, \citenamefont {Kennes}, \citenamefont {Sentef},\ and\ \citenamefont {Yu}}]{cai2023edge}%
  \BibitemOpen
  \bibfield  {author} {\bibinfo {author} {\bibfnamefont {C.}~\bibnamefont {Cai}}, \bibinfo {author} {\bibfnamefont {D.~M.}\ \bibnamefont {Kennes}}, \bibinfo {author} {\bibfnamefont {M.~A.}\ \bibnamefont {Sentef}},\ and\ \bibinfo {author} {\bibfnamefont {T.}~\bibnamefont {Yu}},\ }\bibfield  {title} {\bibinfo {title} {Edge and corner skin effects of chirally coupled magnons characterized by a topological winding tuple},\ }\href@noop {} {\bibfield  {journal} {\bibinfo  {journal} {Physical Review B}\ }\textbf {\bibinfo {volume} {108}},\ \bibinfo {pages} {174421} (\bibinfo {year} {2023})}\BibitemShut {NoStop}%
\bibitem [{\citenamefont {Yu}\ \emph {et~al.}(2024)\citenamefont {Yu}, \citenamefont {Zou}, \citenamefont {Zeng}, \citenamefont {Rao},\ and\ \citenamefont {Xia}}]{yu2024non}%
  \BibitemOpen
  \bibfield  {author} {\bibinfo {author} {\bibfnamefont {T.}~\bibnamefont {Yu}}, \bibinfo {author} {\bibfnamefont {J.}~\bibnamefont {Zou}}, \bibinfo {author} {\bibfnamefont {B.}~\bibnamefont {Zeng}}, \bibinfo {author} {\bibfnamefont {J.}~\bibnamefont {Rao}},\ and\ \bibinfo {author} {\bibfnamefont {K.}~\bibnamefont {Xia}},\ }\bibfield  {title} {\bibinfo {title} {Non-hermitian topological magnonics},\ }\href@noop {} {\bibfield  {journal} {\bibinfo  {journal} {Physics Reports}\ }\textbf {\bibinfo {volume} {1062}},\ \bibinfo {pages} {1} (\bibinfo {year} {2024})}\BibitemShut {NoStop}%
\bibitem [{\citenamefont {Guo}\ \emph {et~al.}(2021)\citenamefont {Guo}, \citenamefont {Liu}, \citenamefont {Zhao}, \citenamefont {Liu},\ and\ \citenamefont {Chen}}]{PhysRevLett.127.116801}%
  \BibitemOpen
  \bibfield  {author} {\bibinfo {author} {\bibfnamefont {C.-X.}\ \bibnamefont {Guo}}, \bibinfo {author} {\bibfnamefont {C.-H.}\ \bibnamefont {Liu}}, \bibinfo {author} {\bibfnamefont {X.-M.}\ \bibnamefont {Zhao}}, \bibinfo {author} {\bibfnamefont {Y.}~\bibnamefont {Liu}},\ and\ \bibinfo {author} {\bibfnamefont {S.}~\bibnamefont {Chen}},\ }\bibfield  {title} {\bibinfo {title} {Exact solution of non-hermitian systems with generalized boundary conditions: Size-dependent boundary effect and fragility of the skin effect},\ }\href {https://doi.org/10.1103/PhysRevLett.127.116801} {\bibfield  {journal} {\bibinfo  {journal} {Phys. Rev. Lett.}\ }\textbf {\bibinfo {volume} {127}},\ \bibinfo {pages} {116801} (\bibinfo {year} {2021})}\BibitemShut {NoStop}%
\bibitem [{\citenamefont {Zhang}\ \emph {et~al.}(2022)\citenamefont {Zhang}, \citenamefont {Zhang}, \citenamefont {Lu},\ and\ \citenamefont {Chen}}]{zhang2022review}%
  \BibitemOpen
  \bibfield  {author} {\bibinfo {author} {\bibfnamefont {X.}~\bibnamefont {Zhang}}, \bibinfo {author} {\bibfnamefont {T.}~\bibnamefont {Zhang}}, \bibinfo {author} {\bibfnamefont {M.-H.}\ \bibnamefont {Lu}},\ and\ \bibinfo {author} {\bibfnamefont {Y.-F.}\ \bibnamefont {Chen}},\ }\bibfield  {title} {\bibinfo {title} {A review on non-hermitian skin effect},\ }\href@noop {} {\bibfield  {journal} {\bibinfo  {journal} {Advances in Physics: X}\ }\textbf {\bibinfo {volume} {7}},\ \bibinfo {pages} {2109431} (\bibinfo {year} {2022})}\BibitemShut {NoStop}%
\bibitem [{\citenamefont {Longhi}(2019)}]{longhi2019probing}%
  \BibitemOpen
  \bibfield  {author} {\bibinfo {author} {\bibfnamefont {S.}~\bibnamefont {Longhi}},\ }\bibfield  {title} {\bibinfo {title} {Probing non-hermitian skin effect and non-bloch phase transitions},\ }\href@noop {} {\bibfield  {journal} {\bibinfo  {journal} {Physical Review Research}\ }\textbf {\bibinfo {volume} {1}},\ \bibinfo {pages} {023013} (\bibinfo {year} {2019})}\BibitemShut {NoStop}%
\bibitem [{\citenamefont {Song}\ \emph {et~al.}(2019)\citenamefont {Song}, \citenamefont {Yao},\ and\ \citenamefont {Wang}}]{song2019non}%
  \BibitemOpen
  \bibfield  {author} {\bibinfo {author} {\bibfnamefont {F.}~\bibnamefont {Song}}, \bibinfo {author} {\bibfnamefont {S.}~\bibnamefont {Yao}},\ and\ \bibinfo {author} {\bibfnamefont {Z.}~\bibnamefont {Wang}},\ }\bibfield  {title} {\bibinfo {title} {Non-hermitian skin effect and chiral damping in open quantum systems},\ }\href@noop {} {\bibfield  {journal} {\bibinfo  {journal} {Physical review letters}\ }\textbf {\bibinfo {volume} {123}},\ \bibinfo {pages} {170401} (\bibinfo {year} {2019})}\BibitemShut {NoStop}%
\bibitem [{\citenamefont {Longhi}(2020)}]{longhi2020unraveling}%
  \BibitemOpen
  \bibfield  {author} {\bibinfo {author} {\bibfnamefont {S.}~\bibnamefont {Longhi}},\ }\bibfield  {title} {\bibinfo {title} {Unraveling the non-hermitian skin effect in dissipative systems},\ }\href@noop {} {\bibfield  {journal} {\bibinfo  {journal} {Physical Review B}\ }\textbf {\bibinfo {volume} {102}},\ \bibinfo {pages} {201103(R)} (\bibinfo {year} {2020})}\BibitemShut {NoStop}%
\bibitem [{\citenamefont {Manna}\ and\ \citenamefont {Roy}(2023)}]{manna2023inner}%
  \BibitemOpen
  \bibfield  {author} {\bibinfo {author} {\bibfnamefont {S.}~\bibnamefont {Manna}}\ and\ \bibinfo {author} {\bibfnamefont {B.}~\bibnamefont {Roy}},\ }\bibfield  {title} {\bibinfo {title} {Inner skin effects on non-hermitian topological fractals},\ }\href@noop {} {\bibfield  {journal} {\bibinfo  {journal} {communications physics}\ }\textbf {\bibinfo {volume} {6}},\ \bibinfo {pages} {10} (\bibinfo {year} {2023})}\BibitemShut {NoStop}%
\bibitem [{\citenamefont {Lin}\ \emph {et~al.}(2023)\citenamefont {Lin}, \citenamefont {Tai}, \citenamefont {Li},\ and\ \citenamefont {Lee}}]{lin2023topological}%
  \BibitemOpen
  \bibfield  {author} {\bibinfo {author} {\bibfnamefont {R.}~\bibnamefont {Lin}}, \bibinfo {author} {\bibfnamefont {T.}~\bibnamefont {Tai}}, \bibinfo {author} {\bibfnamefont {L.}~\bibnamefont {Li}},\ and\ \bibinfo {author} {\bibfnamefont {C.~H.}\ \bibnamefont {Lee}},\ }\bibfield  {title} {\bibinfo {title} {Topological non-hermitian skin effect},\ }\href@noop {} {\bibfield  {journal} {\bibinfo  {journal} {Frontiers of Physics}\ }\textbf {\bibinfo {volume} {18}},\ \bibinfo {pages} {53605} (\bibinfo {year} {2023})}\BibitemShut {NoStop}%
\bibitem [{\citenamefont {Lei}\ \emph {et~al.}(2024)\citenamefont {Lei}, \citenamefont {Lee},\ and\ \citenamefont {Li}}]{lei2024activating}%
  \BibitemOpen
  \bibfield  {author} {\bibinfo {author} {\bibfnamefont {Z.}~\bibnamefont {Lei}}, \bibinfo {author} {\bibfnamefont {C.~H.}\ \bibnamefont {Lee}},\ and\ \bibinfo {author} {\bibfnamefont {L.}~\bibnamefont {Li}},\ }\bibfield  {title} {\bibinfo {title} {Activating non-hermitian skin modes by parity-time symmetry breaking},\ }\href@noop {} {\bibfield  {journal} {\bibinfo  {journal} {Communications Physics}\ }\textbf {\bibinfo {volume} {7}},\ \bibinfo {pages} {100} (\bibinfo {year} {2024})}\BibitemShut {NoStop}%
\bibitem [{\citenamefont {Yang}\ \emph {et~al.}(2022)\citenamefont {Yang}, \citenamefont {Tan}, \citenamefont {Tai}, \citenamefont {Koh}, \citenamefont {Li}, \citenamefont {Longhi},\ and\ \citenamefont {Lee}}]{yang2022designing}%
  \BibitemOpen
  \bibfield  {author} {\bibinfo {author} {\bibfnamefont {R.}~\bibnamefont {Yang}}, \bibinfo {author} {\bibfnamefont {J.~W.}\ \bibnamefont {Tan}}, \bibinfo {author} {\bibfnamefont {T.}~\bibnamefont {Tai}}, \bibinfo {author} {\bibfnamefont {J.~M.}\ \bibnamefont {Koh}}, \bibinfo {author} {\bibfnamefont {L.}~\bibnamefont {Li}}, \bibinfo {author} {\bibfnamefont {S.}~\bibnamefont {Longhi}},\ and\ \bibinfo {author} {\bibfnamefont {C.~H.}\ \bibnamefont {Lee}},\ }\bibfield  {title} {\bibinfo {title} {Designing non-hermitian real spectra through electrostatics},\ }\href@noop {} {\bibfield  {journal} {\bibinfo  {journal} {Science Bulletin}\ }\textbf {\bibinfo {volume} {67}},\ \bibinfo {pages} {1865} (\bibinfo {year} {2022})}\BibitemShut {NoStop}%
\bibitem [{\citenamefont {Ding}\ \emph {et~al.}(2022)\citenamefont {Ding}, \citenamefont {Fang},\ and\ \citenamefont {Ma}}]{ding2022non}%
  \BibitemOpen
  \bibfield  {author} {\bibinfo {author} {\bibfnamefont {K.}~\bibnamefont {Ding}}, \bibinfo {author} {\bibfnamefont {C.}~\bibnamefont {Fang}},\ and\ \bibinfo {author} {\bibfnamefont {G.}~\bibnamefont {Ma}},\ }\bibfield  {title} {\bibinfo {title} {Non-hermitian topology and exceptional-point geometries},\ }\href@noop {} {\bibfield  {journal} {\bibinfo  {journal} {Nature Reviews Physics}\ ,\ \bibinfo {pages} {1}} (\bibinfo {year} {2022})}\BibitemShut {NoStop}%
\bibitem [{\citenamefont {Su}\ \emph {et~al.}(1979)\citenamefont {Su}, \citenamefont {Schrieffer},\ and\ \citenamefont {Heeger}}]{su1979solitons}%
  \BibitemOpen
  \bibfield  {author} {\bibinfo {author} {\bibfnamefont {W.~P.}\ \bibnamefont {Su}}, \bibinfo {author} {\bibfnamefont {J.}~\bibnamefont {Schrieffer}},\ and\ \bibinfo {author} {\bibfnamefont {A.~J.}\ \bibnamefont {Heeger}},\ }\bibfield  {title} {\bibinfo {title} {Solitons in polyacetylene},\ }\href@noop {} {\bibfield  {journal} {\bibinfo  {journal} {Phys. Rev. Lett.}\ }\textbf {\bibinfo {volume} {42}},\ \bibinfo {pages} {1698} (\bibinfo {year} {1979})}\BibitemShut {NoStop}%
\bibitem [{\citenamefont {Heeger}\ \emph {et~al.}(1988)\citenamefont {Heeger}, \citenamefont {Kivelson}, \citenamefont {Schrieffer},\ and\ \citenamefont {Su}}]{heeger1988solitons}%
  \BibitemOpen
  \bibfield  {author} {\bibinfo {author} {\bibfnamefont {A.~J.}\ \bibnamefont {Heeger}}, \bibinfo {author} {\bibfnamefont {S.}~\bibnamefont {Kivelson}}, \bibinfo {author} {\bibfnamefont {J.}~\bibnamefont {Schrieffer}},\ and\ \bibinfo {author} {\bibfnamefont {W.-P.}\ \bibnamefont {Su}},\ }\bibfield  {title} {\bibinfo {title} {Solitons in conducting polymers},\ }\href@noop {} {\bibfield  {journal} {\bibinfo  {journal} {Rev. Mod. Phys.}\ }\textbf {\bibinfo {volume} {60}},\ \bibinfo {pages} {781} (\bibinfo {year} {1988})}\BibitemShut {NoStop}%
\bibitem [{\citenamefont {Longhi}(2018)}]{longhi2018non}%
  \BibitemOpen
  \bibfield  {author} {\bibinfo {author} {\bibfnamefont {S.}~\bibnamefont {Longhi}},\ }\bibfield  {title} {\bibinfo {title} {Non-hermitian gauged topological laser arrays},\ }\href@noop {} {\bibfield  {journal} {\bibinfo  {journal} {Annalen der Physik}\ }\textbf {\bibinfo {volume} {530}},\ \bibinfo {pages} {1800023} (\bibinfo {year} {2018})}\BibitemShut {NoStop}%
\bibitem [{\citenamefont {Yin}\ \emph {et~al.}(2018)\citenamefont {Yin}, \citenamefont {Jiang}, \citenamefont {Li}, \citenamefont {L{\"u}},\ and\ \citenamefont {Chen}}]{yin2018geometrical}%
  \BibitemOpen
  \bibfield  {author} {\bibinfo {author} {\bibfnamefont {C.}~\bibnamefont {Yin}}, \bibinfo {author} {\bibfnamefont {H.}~\bibnamefont {Jiang}}, \bibinfo {author} {\bibfnamefont {L.}~\bibnamefont {Li}}, \bibinfo {author} {\bibfnamefont {R.}~\bibnamefont {L{\"u}}},\ and\ \bibinfo {author} {\bibfnamefont {S.}~\bibnamefont {Chen}},\ }\bibfield  {title} {\bibinfo {title} {Geometrical meaning of winding number and its characterization of topological phases in one-dimensional chiral non-hermitian systems},\ }\href@noop {} {\bibfield  {journal} {\bibinfo  {journal} {Physical Review A}\ }\textbf {\bibinfo {volume} {97}},\ \bibinfo {pages} {052115} (\bibinfo {year} {2018})}\BibitemShut {NoStop}%
\bibitem [{\citenamefont {Rui}\ \emph {et~al.}(2019{\natexlab{b}})\citenamefont {Rui}, \citenamefont {Hirschmann},\ and\ \citenamefont {Schnyder}}]{rui2019pt}%
  \BibitemOpen
  \bibfield  {author} {\bibinfo {author} {\bibfnamefont {W.}~\bibnamefont {Rui}}, \bibinfo {author} {\bibfnamefont {M.~M.}\ \bibnamefont {Hirschmann}},\ and\ \bibinfo {author} {\bibfnamefont {A.~P.}\ \bibnamefont {Schnyder}},\ }\bibfield  {title} {\bibinfo {title} {Pt-symmetric non-hermitian dirac semimetals},\ }\href@noop {} {\bibfield  {journal} {\bibinfo  {journal} {Physical Review B}\ }\textbf {\bibinfo {volume} {100}},\ \bibinfo {pages} {245116} (\bibinfo {year} {2019}{\natexlab{b}})}\BibitemShut {NoStop}%
\bibitem [{\citenamefont {Zhu}\ \emph {et~al.}(2021)\citenamefont {Zhu}, \citenamefont {Teo}, \citenamefont {Li},\ and\ \citenamefont {Gong}}]{zhu2021delocalization}%
  \BibitemOpen
  \bibfield  {author} {\bibinfo {author} {\bibfnamefont {W.}~\bibnamefont {Zhu}}, \bibinfo {author} {\bibfnamefont {W.~X.}\ \bibnamefont {Teo}}, \bibinfo {author} {\bibfnamefont {L.}~\bibnamefont {Li}},\ and\ \bibinfo {author} {\bibfnamefont {J.}~\bibnamefont {Gong}},\ }\bibfield  {title} {\bibinfo {title} {Delocalization of topological edge states},\ }\href@noop {} {\bibfield  {journal} {\bibinfo  {journal} {Physical Review B}\ }\textbf {\bibinfo {volume} {103}},\ \bibinfo {pages} {195414} (\bibinfo {year} {2021})}\BibitemShut {NoStop}%
\bibitem [{\citenamefont {Wang}\ \emph {et~al.}(2022{\natexlab{a}})\citenamefont {Wang}, \citenamefont {Wang},\ and\ \citenamefont {Ma}}]{wang2022non}%
  \BibitemOpen
  \bibfield  {author} {\bibinfo {author} {\bibfnamefont {W.}~\bibnamefont {Wang}}, \bibinfo {author} {\bibfnamefont {X.}~\bibnamefont {Wang}},\ and\ \bibinfo {author} {\bibfnamefont {G.}~\bibnamefont {Ma}},\ }\bibfield  {title} {\bibinfo {title} {Non-hermitian morphing of topological modes},\ }\href@noop {} {\bibfield  {journal} {\bibinfo  {journal} {Nature}\ }\textbf {\bibinfo {volume} {608}},\ \bibinfo {pages} {50} (\bibinfo {year} {2022}{\natexlab{a}})}\BibitemShut {NoStop}%
\bibitem [{\citenamefont {Gao}\ \emph {et~al.}(2020)\citenamefont {Gao}, \citenamefont {Willatzen},\ and\ \citenamefont {Christensen}}]{gao2020anomalous}%
  \BibitemOpen
  \bibfield  {author} {\bibinfo {author} {\bibfnamefont {P.}~\bibnamefont {Gao}}, \bibinfo {author} {\bibfnamefont {M.}~\bibnamefont {Willatzen}},\ and\ \bibinfo {author} {\bibfnamefont {J.}~\bibnamefont {Christensen}},\ }\bibfield  {title} {\bibinfo {title} {Anomalous topological edge states in non-hermitian piezophononic media},\ }\href@noop {} {\bibfield  {journal} {\bibinfo  {journal} {Physical Review Letters}\ }\textbf {\bibinfo {volume} {125}},\ \bibinfo {pages} {206402} (\bibinfo {year} {2020})}\BibitemShut {NoStop}%
\bibitem [{\citenamefont {Wang}\ \emph {et~al.}(2022{\natexlab{b}})\citenamefont {Wang}, \citenamefont {Wang},\ and\ \citenamefont {Ma}}]{wang2022extended}%
  \BibitemOpen
  \bibfield  {author} {\bibinfo {author} {\bibfnamefont {W.}~\bibnamefont {Wang}}, \bibinfo {author} {\bibfnamefont {X.}~\bibnamefont {Wang}},\ and\ \bibinfo {author} {\bibfnamefont {G.}~\bibnamefont {Ma}},\ }\bibfield  {title} {\bibinfo {title} {Extended state in a localized continuum},\ }\href@noop {} {\bibfield  {journal} {\bibinfo  {journal} {Physical Review Letters}\ }\textbf {\bibinfo {volume} {129}},\ \bibinfo {pages} {264301} (\bibinfo {year} {2022}{\natexlab{b}})}\BibitemShut {NoStop}%
\bibitem [{\citenamefont {Cheng}\ \emph {et~al.}(2022)\citenamefont {Cheng}, \citenamefont {Zhang}, \citenamefont {Lu},\ and\ \citenamefont {Chen}}]{cheng2022competition}%
  \BibitemOpen
  \bibfield  {author} {\bibinfo {author} {\bibfnamefont {J.}~\bibnamefont {Cheng}}, \bibinfo {author} {\bibfnamefont {X.}~\bibnamefont {Zhang}}, \bibinfo {author} {\bibfnamefont {M.-H.}\ \bibnamefont {Lu}},\ and\ \bibinfo {author} {\bibfnamefont {Y.-F.}\ \bibnamefont {Chen}},\ }\bibfield  {title} {\bibinfo {title} {Competition between band topology and non-hermiticity},\ }\href@noop {} {\bibfield  {journal} {\bibinfo  {journal} {Physical Review B}\ }\textbf {\bibinfo {volume} {105}},\ \bibinfo {pages} {094103} (\bibinfo {year} {2022})}\BibitemShut {NoStop}%
\bibitem [{\citenamefont {Rui}\ \emph {et~al.}(2023{\natexlab{b}})\citenamefont {Rui}, \citenamefont {Zhao},\ and\ \citenamefont {Wang}}]{rui2023making}%
  \BibitemOpen
  \bibfield  {author} {\bibinfo {author} {\bibfnamefont {W.}~\bibnamefont {Rui}}, \bibinfo {author} {\bibfnamefont {Y.}~\bibnamefont {Zhao}},\ and\ \bibinfo {author} {\bibfnamefont {Z.}~\bibnamefont {Wang}},\ }\bibfield  {title} {\bibinfo {title} {Making topologically trivial non-hermitian systems nontrivial via gauge fields},\ }\href@noop {} {\bibfield  {journal} {\bibinfo  {journal} {Physical Review Letters}\ }\textbf {\bibinfo {volume} {131}},\ \bibinfo {pages} {176402} (\bibinfo {year} {2023}{\natexlab{b}})}\BibitemShut {NoStop}%
\bibitem [{\citenamefont {Eek}\ \emph {et~al.}(2024)\citenamefont {Eek}, \citenamefont {Moustaj}, \citenamefont {R{\"o}ntgen}, \citenamefont {Pagneux}, \citenamefont {Achilleos},\ and\ \citenamefont {Smith}}]{eek2024emergent}%
  \BibitemOpen
  \bibfield  {author} {\bibinfo {author} {\bibfnamefont {L.}~\bibnamefont {Eek}}, \bibinfo {author} {\bibfnamefont {A.}~\bibnamefont {Moustaj}}, \bibinfo {author} {\bibfnamefont {M.}~\bibnamefont {R{\"o}ntgen}}, \bibinfo {author} {\bibfnamefont {V.}~\bibnamefont {Pagneux}}, \bibinfo {author} {\bibfnamefont {V.}~\bibnamefont {Achilleos}},\ and\ \bibinfo {author} {\bibfnamefont {C.~M.}\ \bibnamefont {Smith}},\ }\bibfield  {title} {\bibinfo {title} {Emergent non-hermitian models},\ }\href@noop {} {\bibfield  {journal} {\bibinfo  {journal} {Physical Review B}\ }\textbf {\bibinfo {volume} {109}},\ \bibinfo {pages} {045122} (\bibinfo {year} {2024})}\BibitemShut {NoStop}%
\bibitem [{\citenamefont {Slootman}\ \emph {et~al.}(2024)\citenamefont {Slootman}, \citenamefont {Cherifi}, \citenamefont {Eek}, \citenamefont {Arouca}, \citenamefont {Bergholtz}, \citenamefont {Bourennane},\ and\ \citenamefont {Smith}}]{slootman2024breaking}%
  \BibitemOpen
  \bibfield  {author} {\bibinfo {author} {\bibfnamefont {E.}~\bibnamefont {Slootman}}, \bibinfo {author} {\bibfnamefont {W.}~\bibnamefont {Cherifi}}, \bibinfo {author} {\bibfnamefont {L.}~\bibnamefont {Eek}}, \bibinfo {author} {\bibfnamefont {R.}~\bibnamefont {Arouca}}, \bibinfo {author} {\bibfnamefont {E.}~\bibnamefont {Bergholtz}}, \bibinfo {author} {\bibfnamefont {M.}~\bibnamefont {Bourennane}},\ and\ \bibinfo {author} {\bibfnamefont {C.~M.}\ \bibnamefont {Smith}},\ }\bibfield  {title} {\bibinfo {title} {Breaking and resurgence of symmetry in the non-hermitian su-schrieffer-heeger model in photonic waveguides},\ }\href@noop {} {\bibfield  {journal} {\bibinfo  {journal} {Physical Review Research}\ }\textbf {\bibinfo {volume} {6}},\ \bibinfo {pages} {023140} (\bibinfo {year} {2024})}\BibitemShut {NoStop}%
\bibitem [{\citenamefont {Hou}\ \emph {et~al.}(2022)\citenamefont {Hou}, \citenamefont {Li}, \citenamefont {Chen}, \citenamefont {Liu}, \citenamefont {Yuan}, \citenamefont {Zhang},\ and\ \citenamefont {Ni}}]{hou2022deterministic}%
  \BibitemOpen
  \bibfield  {author} {\bibinfo {author} {\bibfnamefont {C.}~\bibnamefont {Hou}}, \bibinfo {author} {\bibfnamefont {L.}~\bibnamefont {Li}}, \bibinfo {author} {\bibfnamefont {S.}~\bibnamefont {Chen}}, \bibinfo {author} {\bibfnamefont {Y.}~\bibnamefont {Liu}}, \bibinfo {author} {\bibfnamefont {L.}~\bibnamefont {Yuan}}, \bibinfo {author} {\bibfnamefont {Y.}~\bibnamefont {Zhang}},\ and\ \bibinfo {author} {\bibfnamefont {Z.}~\bibnamefont {Ni}},\ }\bibfield  {title} {\bibinfo {title} {Deterministic bulk-boundary correspondences for skin and edge modes in a general two-band non-hermitian system},\ }\href@noop {} {\bibfield  {journal} {\bibinfo  {journal} {Physical Review Research}\ }\textbf {\bibinfo {volume} {4}},\ \bibinfo {pages} {043222} (\bibinfo {year} {2022})}\BibitemShut {NoStop}%
\bibitem [{\citenamefont {Hou}\ \emph {et~al.}(2023)\citenamefont {Hou}, \citenamefont {Li}, \citenamefont {Wu}, \citenamefont {Ruan}, \citenamefont {Chen},\ and\ \citenamefont {Baronio}}]{hou2023topological}%
  \BibitemOpen
  \bibfield  {author} {\bibinfo {author} {\bibfnamefont {C.}~\bibnamefont {Hou}}, \bibinfo {author} {\bibfnamefont {L.}~\bibnamefont {Li}}, \bibinfo {author} {\bibfnamefont {G.}~\bibnamefont {Wu}}, \bibinfo {author} {\bibfnamefont {Y.}~\bibnamefont {Ruan}}, \bibinfo {author} {\bibfnamefont {S.}~\bibnamefont {Chen}},\ and\ \bibinfo {author} {\bibfnamefont {F.}~\bibnamefont {Baronio}},\ }\bibfield  {title} {\bibinfo {title} {Topological edge states in one-dimensional non-hermitian su-schrieffer-heeger systems of finite lattice size: Analytical solutions and exceptional points},\ }\href@noop {} {\bibfield  {journal} {\bibinfo  {journal} {Physical Review B}\ }\textbf {\bibinfo {volume} {108}},\ \bibinfo {pages} {085425} (\bibinfo {year} {2023})}\BibitemShut {NoStop}%
\bibitem [{\citenamefont {Kunst}\ \emph {et~al.}(2018{\natexlab{b}})\citenamefont {Kunst}, \citenamefont {Edvardsson}, \citenamefont {Budich},\ and\ \citenamefont {Bergholtz}}]{PhysRevLett.121.026808}%
  \BibitemOpen
  \bibfield  {author} {\bibinfo {author} {\bibfnamefont {F.~K.}\ \bibnamefont {Kunst}}, \bibinfo {author} {\bibfnamefont {E.}~\bibnamefont {Edvardsson}}, \bibinfo {author} {\bibfnamefont {J.~C.}\ \bibnamefont {Budich}},\ and\ \bibinfo {author} {\bibfnamefont {E.~J.}\ \bibnamefont {Bergholtz}},\ }\bibfield  {title} {\bibinfo {title} {Biorthogonal bulk-boundary correspondence in non-hermitian systems},\ }\href {https://doi.org/10.1103/PhysRevLett.121.026808} {\bibfield  {journal} {\bibinfo  {journal} {Phys. Rev. Lett.}\ }\textbf {\bibinfo {volume} {121}},\ \bibinfo {pages} {026808} (\bibinfo {year} {2018}{\natexlab{b}})}\BibitemShut {NoStop}%
\bibitem [{\citenamefont {Edvardsson}\ \emph {et~al.}(2019)\citenamefont {Edvardsson}, \citenamefont {Kunst},\ and\ \citenamefont {Bergholtz}}]{PhysRevB.99.081302}%
  \BibitemOpen
  \bibfield  {author} {\bibinfo {author} {\bibfnamefont {E.}~\bibnamefont {Edvardsson}}, \bibinfo {author} {\bibfnamefont {F.~K.}\ \bibnamefont {Kunst}},\ and\ \bibinfo {author} {\bibfnamefont {E.~J.}\ \bibnamefont {Bergholtz}},\ }\bibfield  {title} {\bibinfo {title} {Non-hermitian extensions of higher-order topological phases and their biorthogonal bulk-boundary correspondence},\ }\href {https://doi.org/10.1103/PhysRevB.99.081302} {\bibfield  {journal} {\bibinfo  {journal} {Phys. Rev. B}\ }\textbf {\bibinfo {volume} {99}},\ \bibinfo {pages} {081302(R)} (\bibinfo {year} {2019})}\BibitemShut {NoStop}%
\bibitem [{\citenamefont {Bergholtz}\ \emph {et~al.}(2021)\citenamefont {Bergholtz}, \citenamefont {Budich},\ and\ \citenamefont {Kunst}}]{RevModPhys.93.015005}%
  \BibitemOpen
  \bibfield  {author} {\bibinfo {author} {\bibfnamefont {E.~J.}\ \bibnamefont {Bergholtz}}, \bibinfo {author} {\bibfnamefont {J.~C.}\ \bibnamefont {Budich}},\ and\ \bibinfo {author} {\bibfnamefont {F.~K.}\ \bibnamefont {Kunst}},\ }\bibfield  {title} {\bibinfo {title} {Exceptional topology of non-hermitian systems},\ }\href {https://doi.org/10.1103/RevModPhys.93.015005} {\bibfield  {journal} {\bibinfo  {journal} {Rev. Mod. Phys.}\ }\textbf {\bibinfo {volume} {93}},\ \bibinfo {pages} {015005} (\bibinfo {year} {2021})}\BibitemShut {NoStop}%
\bibitem [{\citenamefont {Yi}\ and\ \citenamefont {Yang}(2020)}]{yi2020non}%
  \BibitemOpen
  \bibfield  {author} {\bibinfo {author} {\bibfnamefont {Y.}~\bibnamefont {Yi}}\ and\ \bibinfo {author} {\bibfnamefont {Z.}~\bibnamefont {Yang}},\ }\bibfield  {title} {\bibinfo {title} {Non-hermitian skin modes induced by on-site dissipations and chiral tunneling effect},\ }\href@noop {} {\bibfield  {journal} {\bibinfo  {journal} {Physical Review Letters}\ }\textbf {\bibinfo {volume} {125}},\ \bibinfo {pages} {186802} (\bibinfo {year} {2020})}\BibitemShut {NoStop}%
\bibitem [{\citenamefont {Li}\ \emph {et~al.}(2022)\citenamefont {Li}, \citenamefont {Liang}, \citenamefont {Wang}, \citenamefont {Lu},\ and\ \citenamefont {Liu}}]{li2022gain}%
  \BibitemOpen
  \bibfield  {author} {\bibinfo {author} {\bibfnamefont {Y.}~\bibnamefont {Li}}, \bibinfo {author} {\bibfnamefont {C.}~\bibnamefont {Liang}}, \bibinfo {author} {\bibfnamefont {C.}~\bibnamefont {Wang}}, \bibinfo {author} {\bibfnamefont {C.}~\bibnamefont {Lu}},\ and\ \bibinfo {author} {\bibfnamefont {Y.-C.}\ \bibnamefont {Liu}},\ }\bibfield  {title} {\bibinfo {title} {Gain-loss-induced hybrid skin-topological effect},\ }\href@noop {} {\bibfield  {journal} {\bibinfo  {journal} {Physical Review Letters}\ }\textbf {\bibinfo {volume} {128}},\ \bibinfo {pages} {223903} (\bibinfo {year} {2022})}\BibitemShut {NoStop}%
\bibitem [{\citenamefont {Wu}\ \emph {et~al.}(2022)\citenamefont {Wu}, \citenamefont {Yang}, \citenamefont {Tang}, \citenamefont {Liu},\ and\ \citenamefont {Chen}}]{wu2022flux}%
  \BibitemOpen
  \bibfield  {author} {\bibinfo {author} {\bibfnamefont {C.}~\bibnamefont {Wu}}, \bibinfo {author} {\bibfnamefont {Z.}~\bibnamefont {Yang}}, \bibinfo {author} {\bibfnamefont {J.}~\bibnamefont {Tang}}, \bibinfo {author} {\bibfnamefont {N.}~\bibnamefont {Liu}},\ and\ \bibinfo {author} {\bibfnamefont {G.}~\bibnamefont {Chen}},\ }\bibfield  {title} {\bibinfo {title} {Flux-controlled skin effect and topological transition in a dissipative two-leg ladder model},\ }\href@noop {} {\bibfield  {journal} {\bibinfo  {journal} {Physical Review A}\ }\textbf {\bibinfo {volume} {106}},\ \bibinfo {pages} {062206} (\bibinfo {year} {2022})}\BibitemShut {NoStop}%
\bibitem [{\citenamefont {Xue}\ \emph {et~al.}(2021)\citenamefont {Xue}, \citenamefont {Li}, \citenamefont {Hu}, \citenamefont {Song},\ and\ \citenamefont {Wang}}]{xue2021simple}%
  \BibitemOpen
  \bibfield  {author} {\bibinfo {author} {\bibfnamefont {W.-T.}\ \bibnamefont {Xue}}, \bibinfo {author} {\bibfnamefont {M.-R.}\ \bibnamefont {Li}}, \bibinfo {author} {\bibfnamefont {Y.-M.}\ \bibnamefont {Hu}}, \bibinfo {author} {\bibfnamefont {F.}~\bibnamefont {Song}},\ and\ \bibinfo {author} {\bibfnamefont {Z.}~\bibnamefont {Wang}},\ }\bibfield  {title} {\bibinfo {title} {Simple formulas of directional amplification from non-bloch band theory},\ }\href@noop {} {\bibfield  {journal} {\bibinfo  {journal} {Physical Review B}\ }\textbf {\bibinfo {volume} {103}},\ \bibinfo {pages} {L241408} (\bibinfo {year} {2021})}\BibitemShut {NoStop}%
\bibitem [{\citenamefont {Yokomizo}\ and\ \citenamefont {Murakami}(2020)}]{yokomizo2020non}%
  \BibitemOpen
  \bibfield  {author} {\bibinfo {author} {\bibfnamefont {K.}~\bibnamefont {Yokomizo}}\ and\ \bibinfo {author} {\bibfnamefont {S.}~\bibnamefont {Murakami}},\ }\bibfield  {title} {\bibinfo {title} {Non-bloch band theory and bulk--edge correspondence in non-hermitian systems},\ }\href@noop {} {\bibfield  {journal} {\bibinfo  {journal} {Progress of Theoretical and Experimental Physics}\ }\textbf {\bibinfo {volume} {2020}},\ \bibinfo {pages} {12A102} (\bibinfo {year} {2020})}\BibitemShut {NoStop}%
\bibitem [{\citenamefont {Chiu}\ \emph {et~al.}(2016{\natexlab{b}})\citenamefont {Chiu}, \citenamefont {Teo}, \citenamefont {Schnyder},\ and\ \citenamefont {Ryu}}]{chiu2016classification}%
  \BibitemOpen
  \bibfield  {author} {\bibinfo {author} {\bibfnamefont {C.-K.}\ \bibnamefont {Chiu}}, \bibinfo {author} {\bibfnamefont {J.~C.}\ \bibnamefont {Teo}}, \bibinfo {author} {\bibfnamefont {A.~P.}\ \bibnamefont {Schnyder}},\ and\ \bibinfo {author} {\bibfnamefont {S.}~\bibnamefont {Ryu}},\ }\bibfield  {title} {\bibinfo {title} {Classification of topological quantum matter with symmetries},\ }\href@noop {} {\bibfield  {journal} {\bibinfo  {journal} {Rev. Mod. Phys.}\ }\textbf {\bibinfo {volume} {88}},\ \bibinfo {pages} {035005} (\bibinfo {year} {2016}{\natexlab{b}})}\BibitemShut {NoStop}%
\bibitem [{Sup()}]{SupplementalMaterials}%
  \BibitemOpen
  \href@noop {} {\bibinfo {title} {{See Supplemental Materials at [URL] for non-Hermitian edge states and the equivalence between topological invariant $w_{\scriptscriptstyle GBZ}$ and the global Berry Phase. The Supplemental Materials also contains Refs. [15, 23, 66].}}}\BibitemShut {Stop}%
\bibitem [{\citenamefont {Yu-Min}\ \emph {et~al.}(2021)\citenamefont {Yu-Min}, \citenamefont {Fei},\ and\ \citenamefont {Zhong}}]{yu2021generalized}%
  \BibitemOpen
  \bibfield  {author} {\bibinfo {author} {\bibfnamefont {H.}~\bibnamefont {Yu-Min}}, \bibinfo {author} {\bibfnamefont {S.}~\bibnamefont {Fei}},\ and\ \bibinfo {author} {\bibfnamefont {W.}~\bibnamefont {Zhong}},\ }\bibfield  {title} {\bibinfo {title} {Generalized brillouin zone and non-hermitian band theory},\ }\href@noop {} {\bibfield  {journal} {\bibinfo  {journal} {Acta Physica Sinica}\ }\textbf {\bibinfo {volume} {70}} (\bibinfo {year} {2021})}\BibitemShut {NoStop}%
\bibitem [{\citenamefont {Kawarabayashi}\ and\ \citenamefont {Hatsugai}(2021)}]{kawarabayashi2021bulk}%
  \BibitemOpen
  \bibfield  {author} {\bibinfo {author} {\bibfnamefont {T.}~\bibnamefont {Kawarabayashi}}\ and\ \bibinfo {author} {\bibfnamefont {Y.}~\bibnamefont {Hatsugai}},\ }\bibfield  {title} {\bibinfo {title} {Bulk-edge correspondence with generalized chiral symmetry},\ }\href@noop {} {\bibfield  {journal} {\bibinfo  {journal} {Physical Review B}\ }\textbf {\bibinfo {volume} {103}},\ \bibinfo {pages} {205306} (\bibinfo {year} {2021})}\BibitemShut {NoStop}%
\bibitem [{\citenamefont {Liang}\ and\ \citenamefont {Huang}(2013)}]{liang2013topological}%
  \BibitemOpen
  \bibfield  {author} {\bibinfo {author} {\bibfnamefont {S.-D.}\ \bibnamefont {Liang}}\ and\ \bibinfo {author} {\bibfnamefont {G.-Y.}\ \bibnamefont {Huang}},\ }\bibfield  {title} {\bibinfo {title} {Topological invariance and global berry phase in non-hermitian systems},\ }\href@noop {} {\bibfield  {journal} {\bibinfo  {journal} {Physical Review A—Atomic, Molecular, and Optical Physics}\ }\textbf {\bibinfo {volume} {87}},\ \bibinfo {pages} {012118} (\bibinfo {year} {2013})}\BibitemShut {NoStop}%
\end{thebibliography}%

\end{document}